\documentclass[fleqn,usenatbib]{mnras}

\usepackage[T1]{fontenc}

\DeclareRobustCommand{\VAN}[3]{#2}
\let\VANthebibliography\thebibliography
\def\thebibliography{\DeclareRobustCommand{\VAN}[3]{##3}\VANthebibliography}

\usepackage{graphicx}	
\usepackage{amsmath}	
\usepackage{amssymb}	
\usepackage{newtxtext,newtxmath}

\usepackage{array,multirow}
\usepackage{booktabs}
\usepackage{hyperref}
\usepackage{mathtools}
\usepackage{bm}
\usepackage{gensymb}

\usepackage{units}

\def\app#1#2{%
  \mathrel{%
    \setbox0=\hbox{$#1\sim$}%
    \setbox2=\hbox{%
      \rlap{\hbox{$#1\propto$}}%
      \lower1.1\ht0\box0%
    }%
    \raise0.25\ht2\box2%
  }%
}
\def\approxprop{\mathpalette\app\relax}

\title[Multi-Peaked Light Curves]{Multi-Peaked Non-Thermal Light Curves from Magnetar-Powered Gamma-Ray Bursts}

\author[Omand et al.]{
Conor M. B. Omand \href{https://orcid.org/0000-0002-9646-8710}{\includegraphics[scale=0.5]{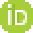}},$^{1}$\thanks{E-mail: c.m.omand@ljmu.ac.uk}
Nikhil Sarin \href{https://orcid.org/0000-0003-2700-1030}{\includegraphics[scale=0.5]{ORCIDiD_icon16x16.eps}},$^{2,3}$ 
and Gavin P. Lamb \href{0000-0001-5169-4143}{\includegraphics[scale=0.5]{ORCIDiD_icon16x16.eps}}$^{1}$ 
\\
$^{1}$Astrophysics Research Institute, Liverpool John Moores University, Liverpool Science Park IC2, 146 Brownlow Hill, Liverpool, UK, L3 5R \\
$^{2}$The Oskar Klein Centre, Department of Physics, Stockholm University, AlbaNova, SE-106 91 Stockholm, Sweden\\
$^{3}$Nordita,
Stockholm University and KTH Royal Institute of Technology
Hannes Alfvéns väg 12, SE-106 91 Stockholm, Sweden\\
}

\date{Accepted XXX. Received YYY; in original form ZZZ}

\pubyear{2024}

\begin{document}
\label{firstpage}
\pagerange{\pageref{firstpage}--\pageref{lastpage}}
\maketitle

\begin{abstract}
Binary neutron star mergers and collapsing massive stars can both create millisecond magnetars.   Such magnetars are candidate engines to power gamma-ray bursts (GRBs).  The non-thermal light curve of the resulting transients can exhibit multiple components, including the GRB afterglow, pulsar wind nebula (PWN), and ejecta afterglow.  We derive the timescales for the peak of each component and show that the PWN is detectable at radio frequencies, dominating the emission for $\sim$ 6 years for supernova/long GRBs (SN/LGRBs) and $\sim$ 100 days for kilonova/short GRBs (KN/SGRBs) at 1 GHz, and $\sim$ 1 year for SN/LGRBs and $\sim$ 15 days for KN/SGRBs at 100 GHz.  The PWN emission has an exponential, frequency-dependent rise to peak that cannot be replicated by an ejecta afterglow.  We show that PWNe in SN/LGRBs can be detected out to $z \sim 0.06$ with current instruments and $z \sim 0.3$ with next-generation instruments and PWNe in KN/SGRBs can be detected out to $z \sim 0.3$ with current instruments and $z \sim 1.5$ with next-generation instruments.  We find that the optimal strategy for detecting PWNe in these systems is a multi-band, high cadence radio follow-up of nearby KN/SGRBs with an x-ray plateau or extended prompt emission from 10 -- 100 days post-burst.
\end{abstract}

\begin{keywords}
stars: magnetars, transients: gamma-ray bursts, transients: neutron star mergers, transients: supernovae, radiation mechanisms: non-thermal
\end{keywords}



\section{Introduction}

Gamma-ray bursts (GRBs) are short flashes of high-energy radiation formed within relativistic jets \citep{Piran2004, Gehrels2012, Zhang2018}.  They are typically divided into two categories based on the duration of their prompt emission, with short GRBs (SGRBs) having a duration of $T_{90} \lesssim 2$ s and long GRBs (LGRBs) having a duration of $T_{90} \gtrsim 2$ s \citep{Kouveliotou1993}, where $T_{90}$ is the timescale over which 90\% of the background-subtracted counts are observed.  The two GRB classes also show differences in their spectra, with SGRBs typically showing harder spectra than LGRBs \citep{Kouveliotou1993}.  This dichotomy supports the idea that these distinct GRB classes arise from different progenitor channels, with SGRBs originating from compact object mergers \citep{Lattimer1976, Blinnikov1984, Eichler1989} and LGRBs originating from the collapse of massive stars \citep{MacFadyen1999, MacFadyen2001}.  There is observational evidence that supports this distinction; LGRBs have been associated with supernovae (SNe) \citep[e.g.,][]{Gendre2013, Nakauchi2013, Levan2014, Cano2017}, and several SGRBs associated with kilonovae (KNe) \citep[e.g.,][]{Tanvir2013, Yang2015, Jin2016, Jin2020, Lamb2019_1608, Fong2021, Zhou2023, Zhu2023}, including the SGRB GRB170817A accompanying the binary neutron star (BNS) merger that produced GW170817 \citep{Abbott2017mm}.  However, two recent LGRBs have been inferred to have originated from BNS mergers due to emission resembling that of a kilonova \citep{Rastinejad2022, Levan2024}, casting some doubt on the robustness of this classification scheme.

Most models for launching relativistic jets that produce GRBs require angular momentum and energy from a compact object -- accretion disk system.  The compact object can either be a black hole or a highly magnetized neutron star, known as a magnetar \citep{Duncan1992}.  For black holes, the energy can be extracted from the black hole spin \citep{Blandford1977} or accretion disk \citep{Blandford1982} via a twisted magnetic field, or neutrino winds from the accretion disk \citep[e.g.][]{Popham1999}.  While magnetars can launch jets via similar processes, they can also launch jets from a magnetorotational mechanism with strong dipolar or toroidal fields \citep{Mosta2020, Bugli2021}, or from strong propeller-driven outflows \citep{Illarionov1975, Lovelace1999, Romanova2005}.

When the relativistic jet sweeps up material and starts to decelerate, a broadband synchrotron and inverse Compton GRB afterglow is emitted \citep{Paczynski1993, Sari1998, Piran2004, Zhang2004, Zhang2006, MAGICCollaboration2019}.  The temporal lightcurve of this afterglow will be dominated by the jet core for on-axis observers and will depend on the wider structure of the jet for off-axis observers \citep{Granot2002, Rossi2002}.  While the afterglow emission is mostly predicted to rise and fall as power laws \citep{Sari1998}, several other phenomena have been detected in the afterglow emission.  One of these phenomena are X-ray plateaus, where the flux of the X-ray afterglow transitions from an initial steep decay, to a very shallow decay or plateau, back to a steep decay \citep{Nousek2006, O'Brien2006, Rowlinson2013, Dainotti2017}.  This has been suggested to be due to energy injection from a central magnetar \citep{Troja2007, Rowlinson2013, Gompertz2014, Stratta2018, Sarin2019}.  Another GRB phenomenon potentially linked to a magnetar central engine is extended emission \citep{Metzger2008, Bucciantini2012, Gompertz2013, Gompertz2014, Gibson2017}, where the prompt GRB emission is followed by a lower luminosity signal that can be several times longer than the prompt emission and extends to softer energies~\citep{Norris2006, Bostanci2013, Kaneko2015}.

The wind from the spin-down of a rotating neutron star will also produce a pulsar wind nebula (PWN) on the interior of the ejecta that is a broadband synchrotron and inverse Compton emitter.  The thermalization of PWN emission in the ejecta is thought to power a number of transients, including superluminous supernovae (SLSNe) \citep{Nicholl2017, Kangas2022, West2023, Gomez2024, Gkini2024} and some broad-line SN-Ic (SNe Ic-BL) \citep{Mazzali2014, Greiner2015, Wang2017, Omand2024}.  The PWN can also be detected in radio \citep{Murase2015, Omand2018, Eftekhari2019, Law2019, Mondal2020, Eftekhari2021}.  It can also cause an infrared excess by heating up dust in the ejecta \citep{Omand2019, Chen2021, Sun2022}, change the spectrum by ionizing the ejecta \citep{Chevalier1992, Jerkstrand2017, Omand2023, Dessart2024}, and cause polarization signals by expanding the ejecta asymmetrically \citep{Inserra2016, Saito2020, Poidevin2022, Pursiainen2022, Poidevin2023, Pursiainen2023}.  The PWN can also cause instabilities in the ejecta \citep{Chen2016, Suzuki2017, Suzuki2021} that results in a filamentary ejecta structure similar to the Crab Nebula \citep{Clark1983, Bietenholz1991, Temim2006, Omand2024Crab}.

The ejecta from a transient accompanying a GRB, a supernova or a kilonova, can produce its own afterglow once it sweeps up enough mass to decelerate the ejecta.  X-ray observations of GRB170817A have shown tentative evidence for a re-brightening of the afterglow at late times potentially consistent with this scenario \citep{Hajela2022, Troja2022}.  
A few supernovae have been detected in radio at late times \citep[e.g.][]{vanDyk1994, Gaensler1997, Corsi2022, Kool2023, Margutti2023} and X-ray \citep[e.g.][]{Chevalier1994, Dwarkadas2014, Bochenek2018}, and some at much earlier times when dense circumstellar material (CSM) surrounds the progenitor \citep[e.g.][]{Smith2008, Fransson2014, Drout2016}.  

Magnetar-driven GRB systems should produce non-thermal emission from a GRB afterglow, PWN, and SN/KN afterglow (we refer to this as an ejecta afterglow), as well as thermal emission from the magnetar-heated ejecta.  The timescale for these three non-thermal mechanisms can be significantly different, leading to the potential of detecting all three components separately. The GRB and ejecta afterglow both arise when the jet and ejecta sweep up an amount of mass comparable to their own mass divided by their Lorentz factor \citep{Rees1992, Sari1995}, but the jet is much faster and less massive, leading to the GRB afterglow peaking much earlier.  The peak of the PWN emission is set by the opacity of the ejecta, which can vary greatly across bands due to the different absorption processes involved at different energies.  However, for most energies, the PWN will peak sometime in between the GRB afterglow and ejecta afterglow, leading to a distinct third component.  Previous modeling studies of magnetar-driven GRBs or supernova/GRBs have not considered all three components \citep[e.g.][]{Murase2018, Sarin2022, Kusafuka2025}, and previous observations searching for multiple components have only led to upper limits \citep[e.g. ][]{Schroeder2020}.

Here, we examine the emission for each component and show the multi-peaked non-thermal light curves that can result from such systems.  In Section \ref{sec:timescales}, we derive the scaling relations and fiducial timescales for the three components.  In Section \ref{sec:mod}, we show the resulting light curves in radio and X-ray, and show that the magnetar signal can not be reproduced by an ejecta afterglow.  In Section \ref{sec:disc}, we discuss strategies for observation, caveats and uncertainties of our approach, and analyze a few notable GRBs.  Finally, in Section \ref{sec:sum}, we summarize our findings.

\section{Analytic Timescale Estimates} \label{sec:timescales}

Here, we estimate the timescales of the peaks of each non-thermal component using simple analytic scalings at radio and X-ray frequencies.  The fiducial timescales for each component in both the kilonova and supernova scenarios are summarized in Table \ref{tbl:timescales}.

\subsection{GRB Afterglow}

We estimate the peak timescale for an on-axis afterglow using the standard closure relations \citep{Sari1998}.  We assume a fully adiabatic shock and that the afterglow is in the slow-cooling regime.  We also assume that the density of the ambient medium is constant.  The timescales where the cooling frequency $\nu_c$ and the frequency at the minimum electron acceleration $\nu_m$ cross the observed frequency $\nu$ are
\begin{align}
    t_c \approx & 7.3 \left(\frac{\epsilon_B}{0.01}\right)^{-3} \left(\frac{E_{\rm jet}}{10^{52} \text{ erg}}\right)^{-1} \left(\frac{n_{\rm CSM}}{1 \text{ cm}^{-3}}\right)^{-2} \left(\frac{\nu}{10^{6} \text{ GHz}}\right)^{-2} \text{ days} , \label{eqn:tc} \\
    t_m \approx & 10 \left(\frac{\epsilon_B}{0.01}\right)^{1/3} \left(\frac{\epsilon_e}{0.1}\right)^{4/3} \left(\frac{E_{\rm jet}}{10^{52} \text{ erg}}\right)^{1/3} \left(\frac{\nu}{10^{6} \text{ GHz}}\right)^{-2/3} \text{ minutes} , \label{eqn:tm}
\end{align}
where $\epsilon_B$ and $\epsilon_e$ are the fractions of shock energy that go into magnetic field and leptons, respectively, $E_{\rm jet}$ is the energy of the jet, and $n_{\rm CSM}$ is the number density of the ambient medium.  Whether the peak of the light curve is set by $t_c$ or $t_m$ is determined by which timescale is shorter, which can be determined by the critical frequency
\begin{equation}
    \nu_{\rm crit} \approx 1.8 \times 10^8 \left(\frac{\epsilon_B}{0.01}\right)^{-5/2} \left(\frac{\epsilon_e}{0.1}\right)^{-1} \left(\frac{E_{\rm jet}}{10^{52} \text{ erg}}\right)^{-1} \left(\frac{n_{\rm CSM}}{1 \text{ cm}^{-3}}\right)^{-3/2}  \text{ GHz.}
    \label{eqn:nuo_grb}
\end{equation}
For $\nu > \nu_{\rm crit}$ (the high-frequency light curve), the peak occurs at $t_c$, while for $\nu < \nu_{\rm crit}$ (the low-frequency light curve), the peak occurs at $t_m$.  $\nu = \nu_{\rm crit}$ at the critical timescale is

\begin{equation}
    t_{\rm crit} \approx 18  \left(\frac{\epsilon_B}{0.01}\right)^2 \left(\frac{\epsilon_e}{0.1}\right)^2 \left(\frac{E_{\rm jet}}{10^{52} \text{ erg}}\right) \left(\frac{n_{\rm CSM}}{1 \text{ cm}^{-3}}\right)  \text{ seconds.}
    \label{eqn:to_grb}
\end{equation}
This timescale is much earlier than the timescales we observe, and justifies our assumption of slow cooling.

If these timescales are shorter than the jet deceleration timescale \citep{Blandford1976, Blandford1977mckee, Meszaros2006}

\begin{equation}
  t_{\rm dec, jet} =  100 \left(\frac{E_{\rm jet}}{10^{52} \text{ erg}}\right)^{1/3} \left(\frac{n_{\rm CSM}}{1 \text{ cm}^{-3}}\right)^{-1/3} \left(\frac{\gamma_0}{100}\right)^{-8/3} \text{ s,}
  \label{eqn:tdecjet}
\end{equation}
where $\gamma_0$ is the initial Lorentz factor of the jet, then the light curve will peak at the deceleration time.  For a typical observed LGRB jet with $\epsilon_e \sim 0.1$, $\epsilon_B \sim 0.01$, $E_{\rm jet} \sim 10^{52}$ erg, $\gamma_0 \sim 100$, and $n_{\rm CSM} \sim 1$ cm$^{-3}$ \citep{Wang2015lgrb, Atteia2017}, the critical frequency is $\nu_{\rm crit} \sim 2 \times 10^{8} \text{ GHz } \approx 1$ keV, and the peak timescales in the radio bands are $\sim$ 70 days at 1 GHz and around 3 days at 100 GHz, while in the X-ray bands they are set by the deceleration time, $\sim$ 100 s.  For a typical observed SGRB, with the same $\epsilon_e$, $\epsilon_B$, $\gamma_0$, and $n_{\rm CSM}$, but a lower jet energy of $E_{\rm jet} \sim 10^{50}$ erg \citep{Fong2015}, the critical frequency is $\nu_{\rm crit} \sim 2 \times 10^{11} \text{ GHz } \approx 100$ keV, and the peak timescales in the radio bands are $\sim$ 7 days at 1 GHz and around 8 hours at 100 GHz, while in the X-ray bands they are set by the deceleration time, $\sim$ 20 s.

For an afterglow viewed from off-axis, the peak will be delayed compared to the same afterglow viewed on-axis.  
For $\theta_{\rm obs} \gg \theta_{\rm jet}$, the peak timescale will change $\propto$ $\theta_{\rm obs}^{8/3}$, regardless of the angular structure of the jet\footnote{This applies as long as the jet has a core-dominated structure.} \citep{Nakar2002, Lamb2017, Xie2018, Ioka2018}.

\subsection{PWN}

We estimate the peak timescale for a PWN using the relations derived by the model in Appendix \ref{sec:pwn}. Due to the PWN emission being produced shortly after the explosion from the inside of the ejecta, the timescale for the observed peak is set by when the optical depth of the ejecta in a specific band drops to unity.  For radio emission, the key process is free-free absorption, although synchrotron self-absorption can be key in certain regions of parameter space; for soft X-rays, it is photoelectric absorption; and for hard X-rays, it is Compton scattering.

Taking Equations \ref{eqn:tauff}, \ref{eqn:kpe} and \ref{eqn:kcomp} for optical depth and solving for the timescales where $\tau = 1$ gives

\begin{align}
     t_{\rm esc,ff} \approx {}& 130 \left(\frac{M_{\rm ej}}{M_\odot}\right)^{2/5} \left(\frac{Y_{\rm fe}}{0.05}\right)^{2/5}  \left(\frac{\bar{Z}}{10}\right)^{2/5} \nonumber \\ {}& \hspace{1cm} \left(\frac{v_{\rm ej}}{10^5 \text{ km s}^{-1}}\right)^{-1}  \left(\frac{\nu}{10 \text{ GHz}}\right)^{-0.42}\text{ days, }          
     \label{eqn:tesc_ff} \\
    t_{\rm esc,pe} \approx {}& 80 \left(\frac{M_{\rm ej}}{M_\odot}\right)^{1/2} \left(\frac{\bar{Z}}{10}\right)   \left(\frac{v_{\rm ej}}{10^5 \text{ km s}^{-1}}\right)^{-1} \nonumber  \\ {}& \hspace{3.425cm}
    \left(\frac{h\nu}{10 \text{ keV}}\right)^{-3/2}\text{ days,}
    \label{eqn:tesc_pe} \\
    t_{\rm esc,comp} \approx {}& 10 \left(\frac{M_{\rm ej}}{M_\odot}\right)^{1/2} \left(\frac{Y_{\rm e}}{0.5}\right)^{1/2} \left(\frac{v_{\rm ej}}{10^5 \text{ km s}^{-1}}\right)^{-1}  \nonumber  \\ {}& \hspace{3.5cm}   
    \left(\frac{\sigma_{\rm KN}(\nu)}{\sigma_{\rm T}}\right)^{1/2}\text{ days,}
\end{align}
where $Y_{\rm fe}$ is the free electron fraction (defined in Equation \ref{eqn:Yfe}) within the ejecta, $\bar{Z}$ is the average atomic number of the ejected material, $v_{\rm ej}$ is the  ejecta velocity, $Y_e$ is the electron fraction within the ejecta, and $\sigma_{\rm KN}$ and $\sigma_T$ are the Klein-Nishina and Thompson cross-sections, respectively.

For a fiducial GRB-SN with parameters chosen to be broadly consistent with observations \citep{Taddia2019, Srinivasaragavan2024}, the ejecta mass and velocity will be around 5 $M_\odot$ and 20 000 km s$^{-1}$, giving a total kinetic energy of $\sim$ 10$^{52}$ erg.  The ejecta is assumed to consist of singly ionized oxygen, which has $Y_{\rm e} \sim 0.5$, $\bar{Z} \sim 8$, and $Y_{\rm fe} \sim 1/16$.  The assumption of oxygen ejecta is motivated by previous nucleosyntheis studies on GRB-SNe \citep[e.g.][]{Maeda2002} and the assumption of single-ionization is motivated by studies examining the ionization state of the ejecta in magnetar-driven supernovae \citep{Margalit2018, Omand2023}. Given these parameters, the peak timescale is $\sim$ 10 years at 1 GHz, $\sim$ 500 days at 100 GHz, $\sim$ 60 years at 1 keV, and $\sim$ 100 days at 100 keV.

For our fiducial magnetar-driven kilonova, motivated by previous modeling efforts and numerical simulations \citep{Yu2013, Siegel2017, Murase2018, Metzger2019, Margalit2019, Sarin2022, Ai2024}, the ejecta mass and velocity will be around 0.1 $M_\odot$ and 0.5$c$, giving a total kinetic energy of $\sim 2 \times 10^{52}$ erg.  The ejecta is assumed to have $Y_{\rm e} \sim 0.4$, $\bar{Z} \sim 40$, and $Y_{\rm fe} \sim 0.02$, which is broadly consistent with r-process nucleosynthesis calculations \citep{Foucart2016, Roberts2017, Vlasov2017}.  Given these parameters, the peak timescale is $\sim$ 100 days at 1 GHz, $\sim$ 15 days at 100 GHz, $\sim$ 6 years at 1 keV, and $\sim$ 2 days at 100 keV.

\subsection{Ejecta Afterglow}

The emission from the ejecta-CSM interaction will peak around the time when the ejecta sweeps up a mass comparable to itself, known as the ejecta deceleration timescale.  Following \citet{Nakar2011} \citep[see also][]{Hotokezaka2016}, a spherical outflow with energy $E_{\rm ej}$ and velocity $c\beta_0$ propagating into a medium of constant density $n_{\rm CSM}$ will begin to decelerate at a radius of 
\begin{equation}
    R_{\rm dec} \approx 10^{18} \left(\frac{E_{\rm ej}}{10^{52} \text{ erg}}\right)^{1/3} \left(\frac{n_{\rm CSM}}{1 \text{ cm}^{-3}}\right)^{-1/3} \beta_0^{-2/3} \text{ cm}
    \label{eqn:rdec}
\end{equation}
on a timescale of 
\begin{equation}
    t_{\rm dec, ejecta} = \frac{R_{\rm dec}}{c\beta_0} \approx 300 \left(\frac{E_{\rm ej}}{10^{52} \text{ erg}}\right)^{1/3} \left(\frac{n_{\rm CSM}}{1 \text{ cm}^{-3}}\right)^{-1/3} \beta_0^{-5/3} \text{ days}.
    \label{eqn:tdec}
\end{equation}
$E_{\rm ej}$ is calculated from the initial kinetic energy $E_{\rm K}$ from the explosion/merger and the contribution from the magnetar, which depends of the spin-down luminosity and timescale as well as the ejecta mass (for further discussion, see \citet{Suzuki2021} and \citet{Omand2024}).  The peak timescale can be longer if the observed frequency is below either the self-absorption frequency $\nu_{\rm ssa}$ or the synchrotron frequency at the minimum Lorentz factor $\nu_m$ at the deceleration timescale.  These frequencies have values
\begin{align}
    \nu_{\rm ssa} \approx & 1 \left(\frac{E_{\rm ej}}{10^{49} \text{ erg}}\right)^{\frac{2}{3(4+p)}} \left(\frac{n_{\rm CSM}}{1 \text{ cm}^{-3}}\right)^{\frac{14+3p}{6(4+p)}} \nonumber \\
    &\left(\frac{\epsilon_B}{0.1}\right)^{\frac{2+p}{2(4+p)}} \left(\frac{\epsilon_e}{0.1}\right)^{\frac{2(p-1)}{4+p}} \beta_0^{\frac{15p-10}{3(4+p)}} \hspace{0.33cm} \text{ GHz} \label{eqn:nussa} \\
    \nu_m \approx & 1 \left(\frac{n_{\rm CSM}}{1 \text{ cm}^{-3}}\right)^{1/2} \left(\frac{\epsilon_B}{0.1}\right)^{1/2} \left(\frac{\epsilon_e}{0.1}\right)^{2} \beta_0^{5} \text{ GHz} \label{eqn:num}
\end{align}
at $t_{\rm dec}$, where $p \sim 2.5$ is the index of the accelerated electrons $dN/d\gamma \propto \gamma^{-p}$.  For a fiducial GRB supernova ($E_{\rm ej} \sim 10^{52}$ erg, $\beta_0 \sim 0.06$, $n_{\rm CSM} \sim 1$ cm$^{-3}$, $\epsilon_B$ = $\epsilon_e \sim $ 0.1) and for ejecta from a magnetar-driven kilonova ($E_{\rm ej} \sim 10^{52}$ erg, $\beta_0 \sim 0.5$, $n_{\rm CSM} \sim 1$ cm$^{-3}$, $\epsilon_B$ = $\epsilon_e \sim $ 0.1) \citep{Nakar2011, Sarin2022}, $\nu_{\rm ssa}$ and $\nu_m$ are both below 1 GHz, and thus the light curve in all bands above 1 GHz will peak at the deceleration timescale, which is $\sim$ 3 years for the kilonova and $\sim$ 80 years for the supernova.

\begin{table*}
    \centering
    \begin{tabular}{|c|c|cccc|} \hline
       Transient & Component & 1 GHz & 100 GHz & 1 keV & 100 keV \\ \hline
        & GRB Afterglow & 70 days & 3 days & 100 s & 100 s \\
       SN/LGRB & PWN & 10 yr & 500 days & 60 yr & 100 days \\
       & Ejecta Afterglow & 80 yr & 80 yr & 80 yr & 80 yr \\ \hline
      & GRB Afterglow & 7 days & 8 hours & 20 s & 20 s \\
       KN/SGRB & PWN & 100 days & 15 days & 6 yr & 2 days \\
       & Ejecta Afterglow & 3 yr & 3 yr & 3 yr & 3 yr \\ \hline
    \end{tabular}
    \caption{The peak timescale of each non-thermal component at various radio and X-ray bands for our fiducial parameters.  The parameters used for estimating the timescales and the scaling relations for each value can be found in their respective sections.}
    \label{tbl:timescales}
\end{table*}

\section{Modeling} \label{sec:mod}

\subsection{Fiducial Light Curves} \label{sec:typical}

\begin{figure*}
\newcolumntype{D}{>{\centering\arraybackslash} m{4cm}}
\noindent
\makebox[\textwidth]{
\begin{tabular}{m{1.1cm} DDDD}
& \textbf{1 GHz} & \textbf{100 GHz}  & \textbf{1 keV} & \textbf{100 keV} \\
\textbf{SN/LGRB (on-axis)} &
\includegraphics[width=1.08\linewidth]{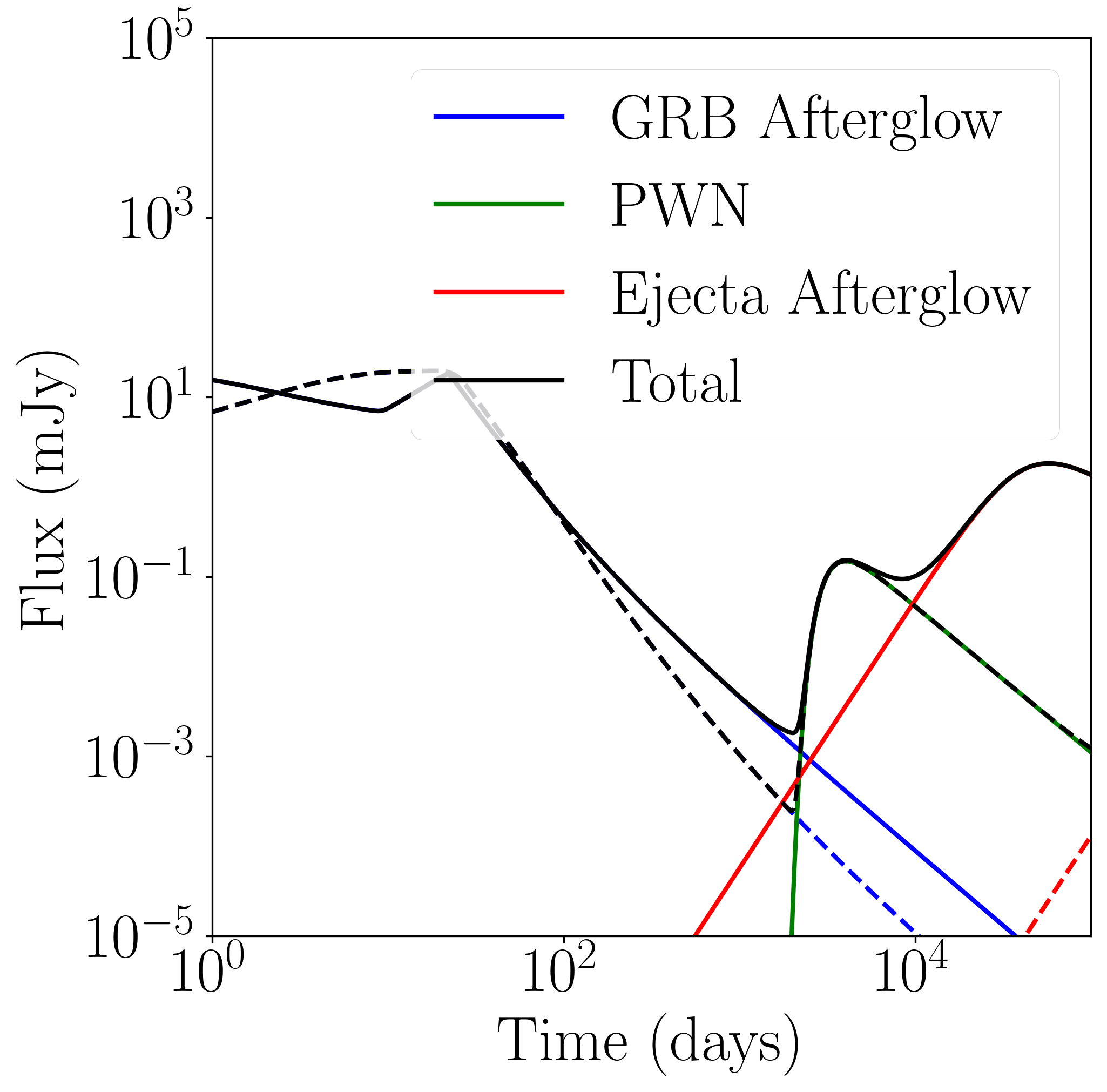}&
\includegraphics[width=1.08\linewidth]{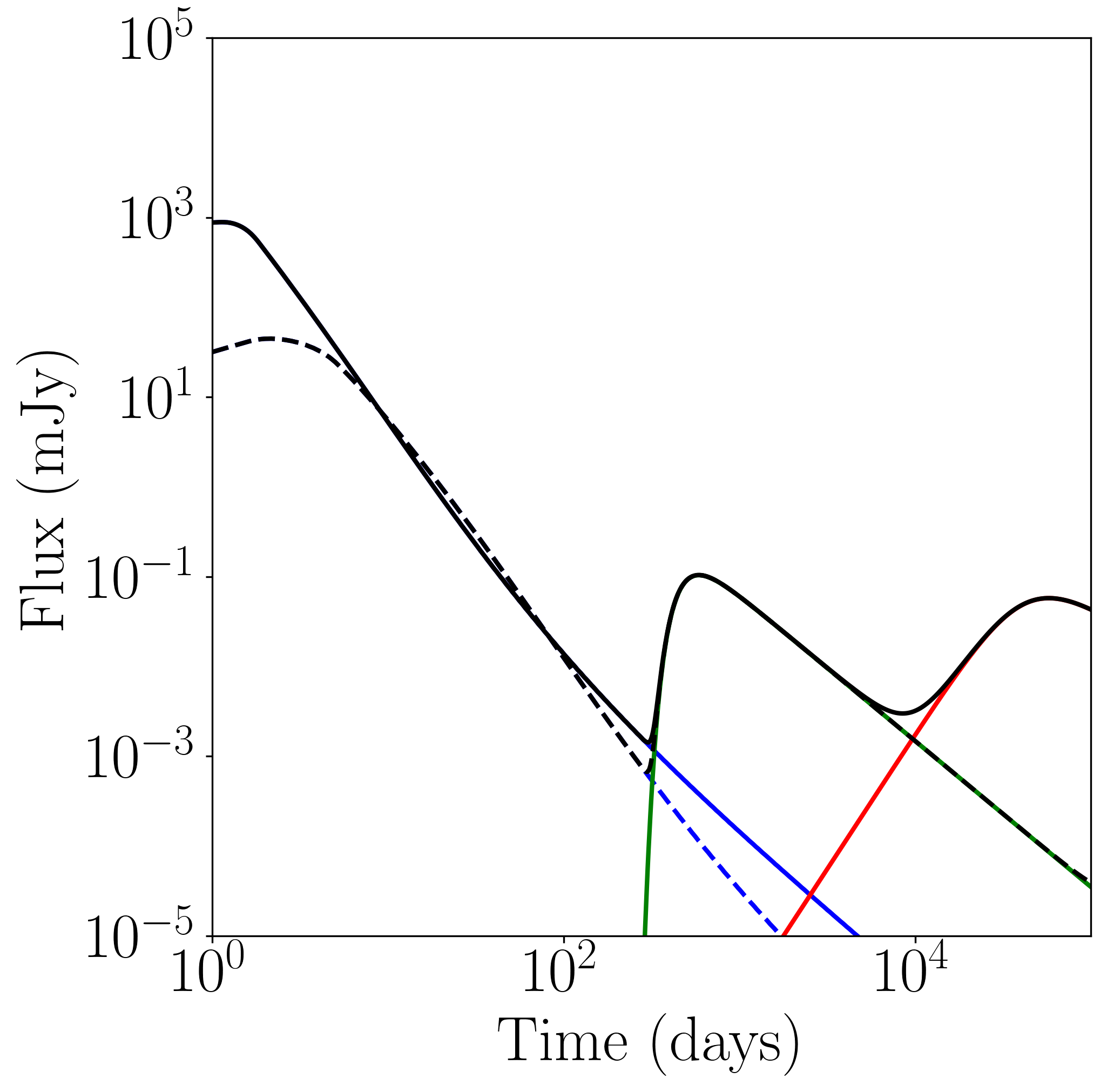}&
\includegraphics[width=1.08\linewidth]{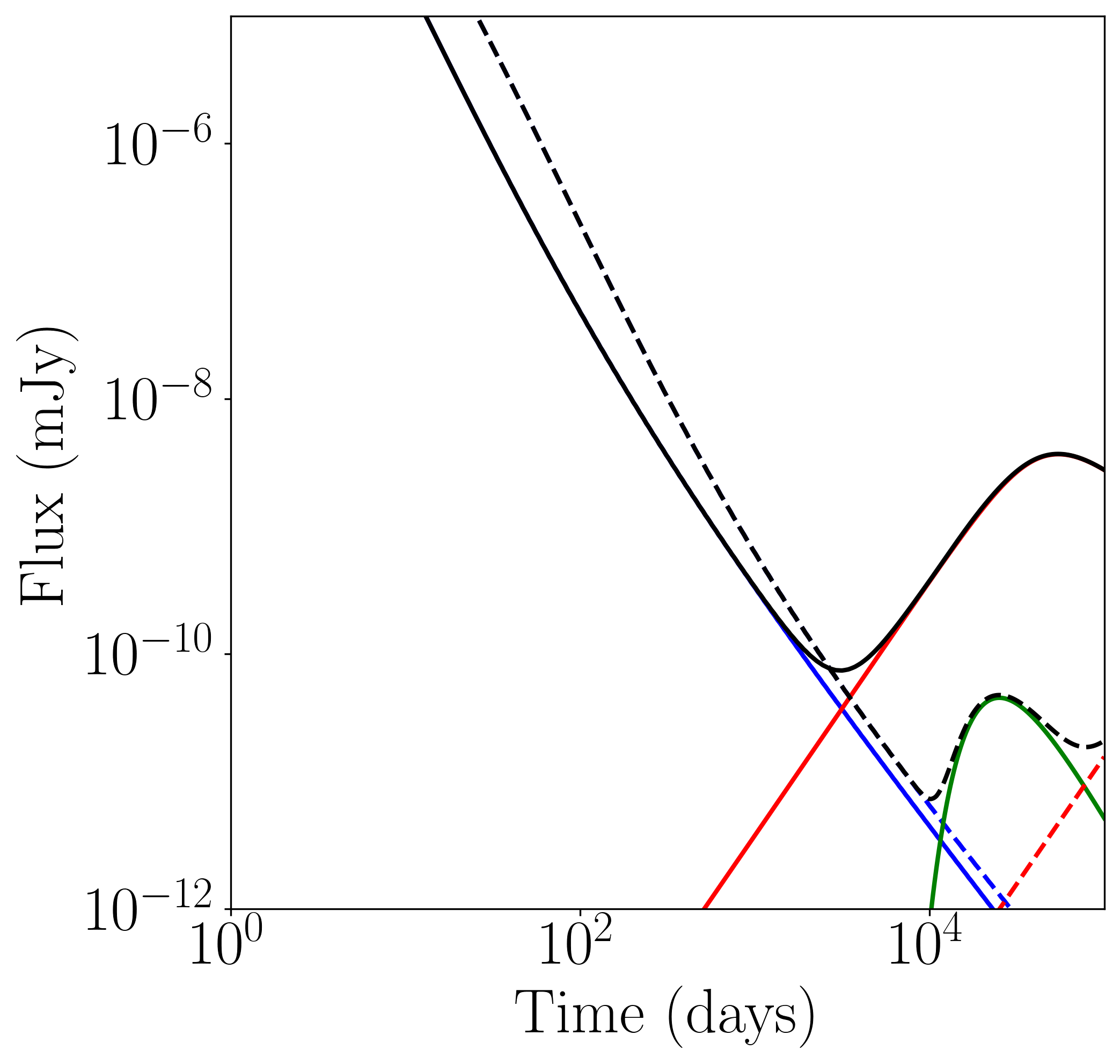}&
\includegraphics[width=1.08\linewidth]{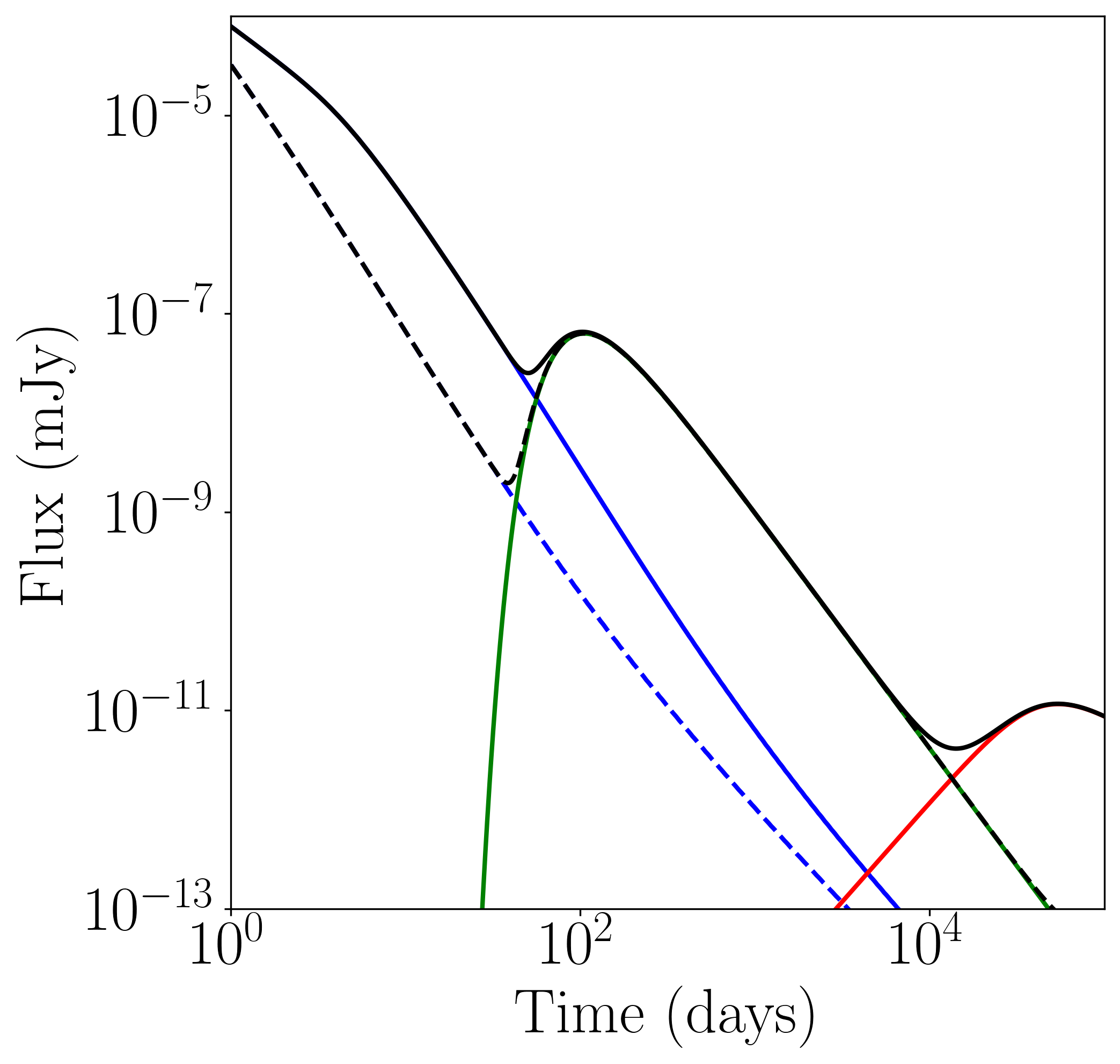}\\[-1.05ex]
\textbf{SN/LGRB (off-axis)}&
\includegraphics[width=1.08\linewidth]{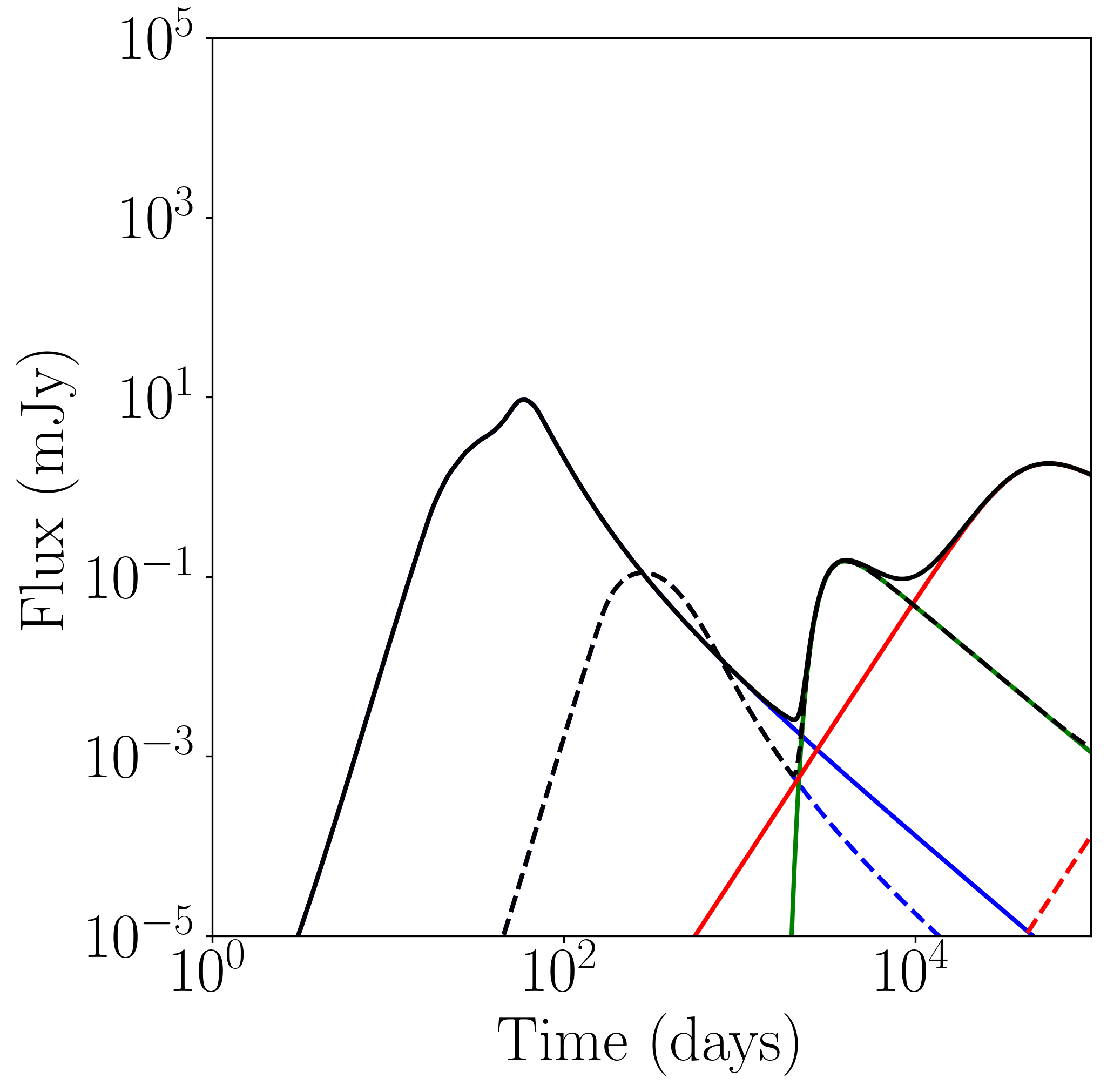}&
\includegraphics[width=1.08\linewidth]{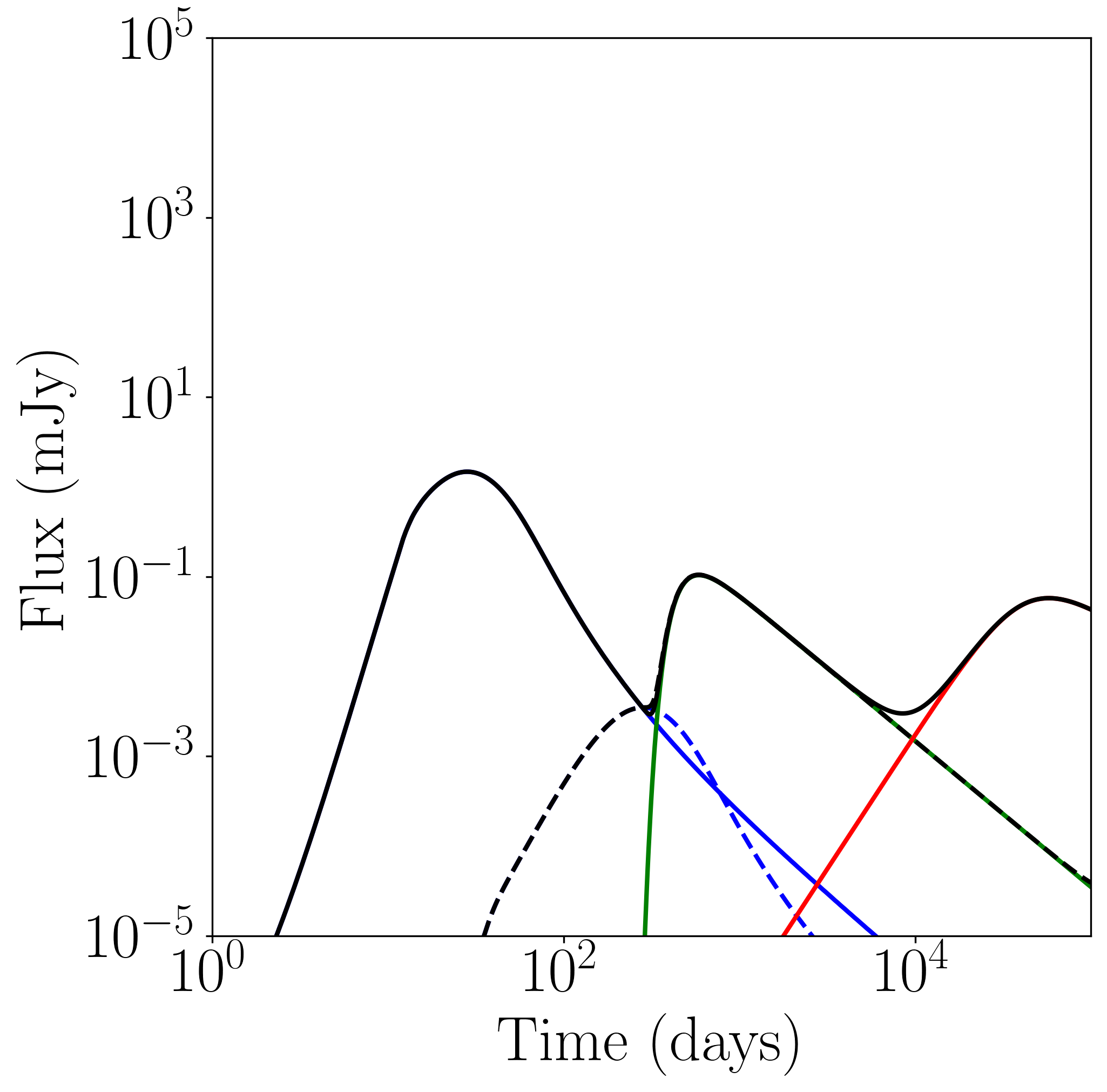}&
\includegraphics[width=1.08\linewidth]{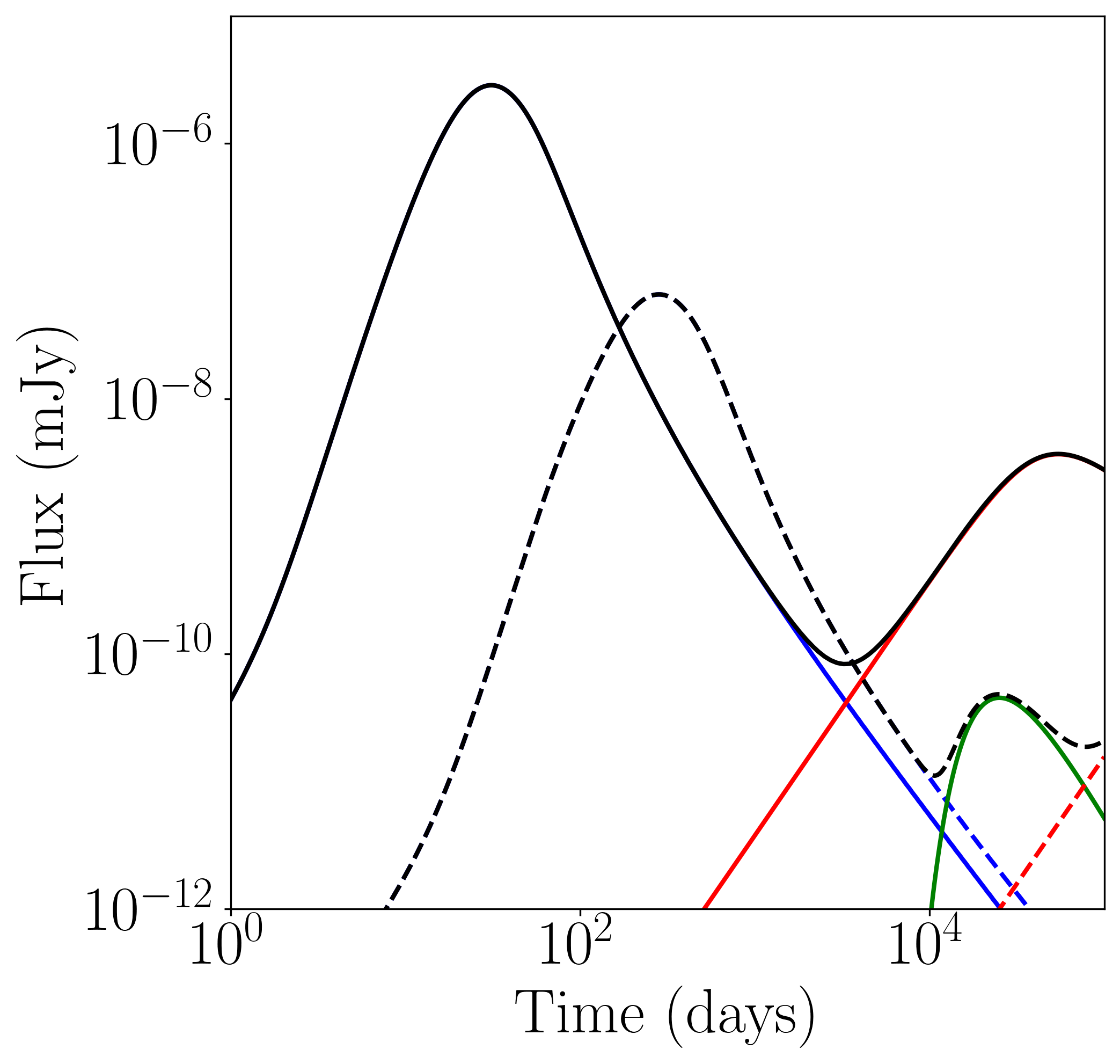}&
\includegraphics[width=1.08\linewidth]{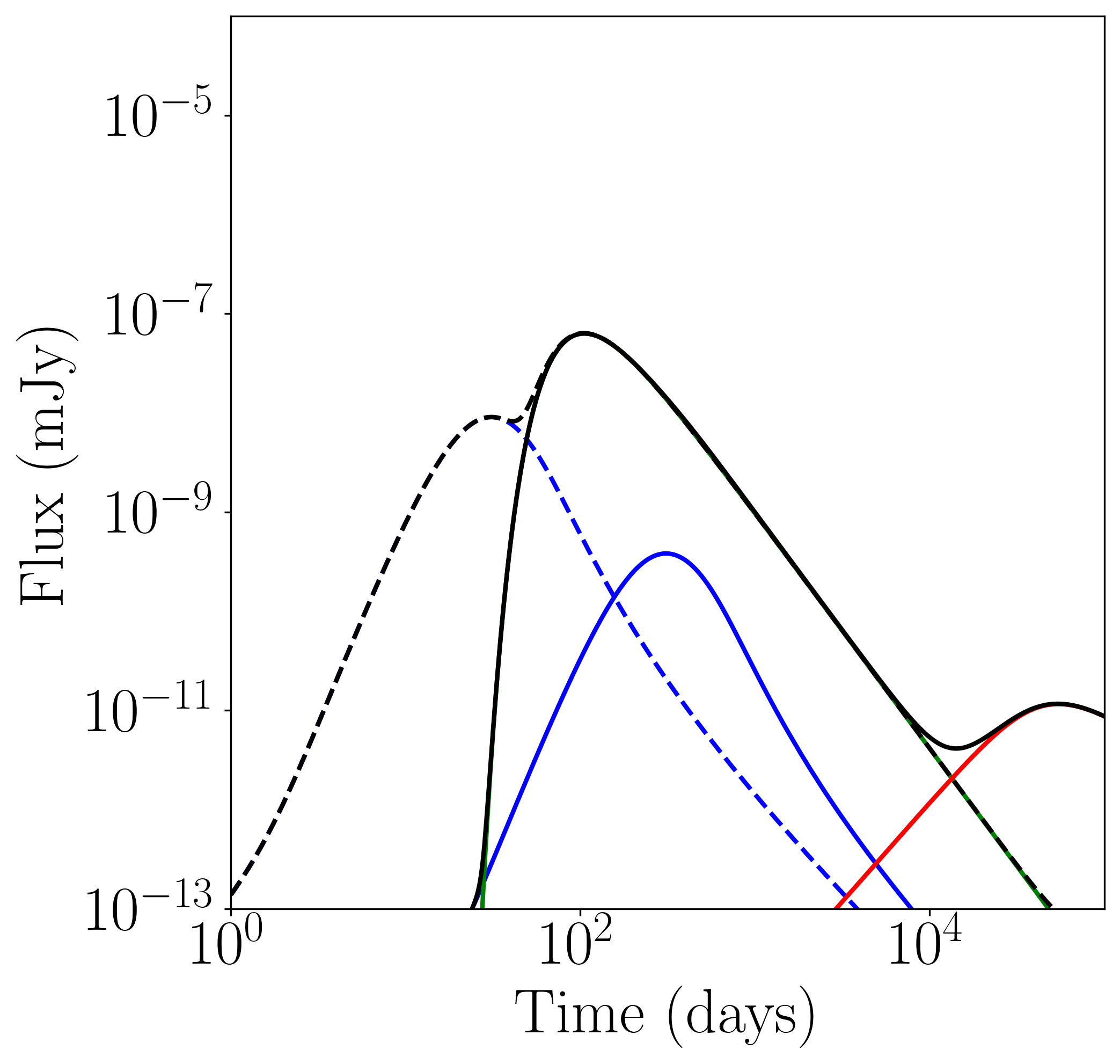}\\[-1.05ex]
\textbf{KN/SGRB (on-axis)} &
\includegraphics[width=1.08\linewidth]{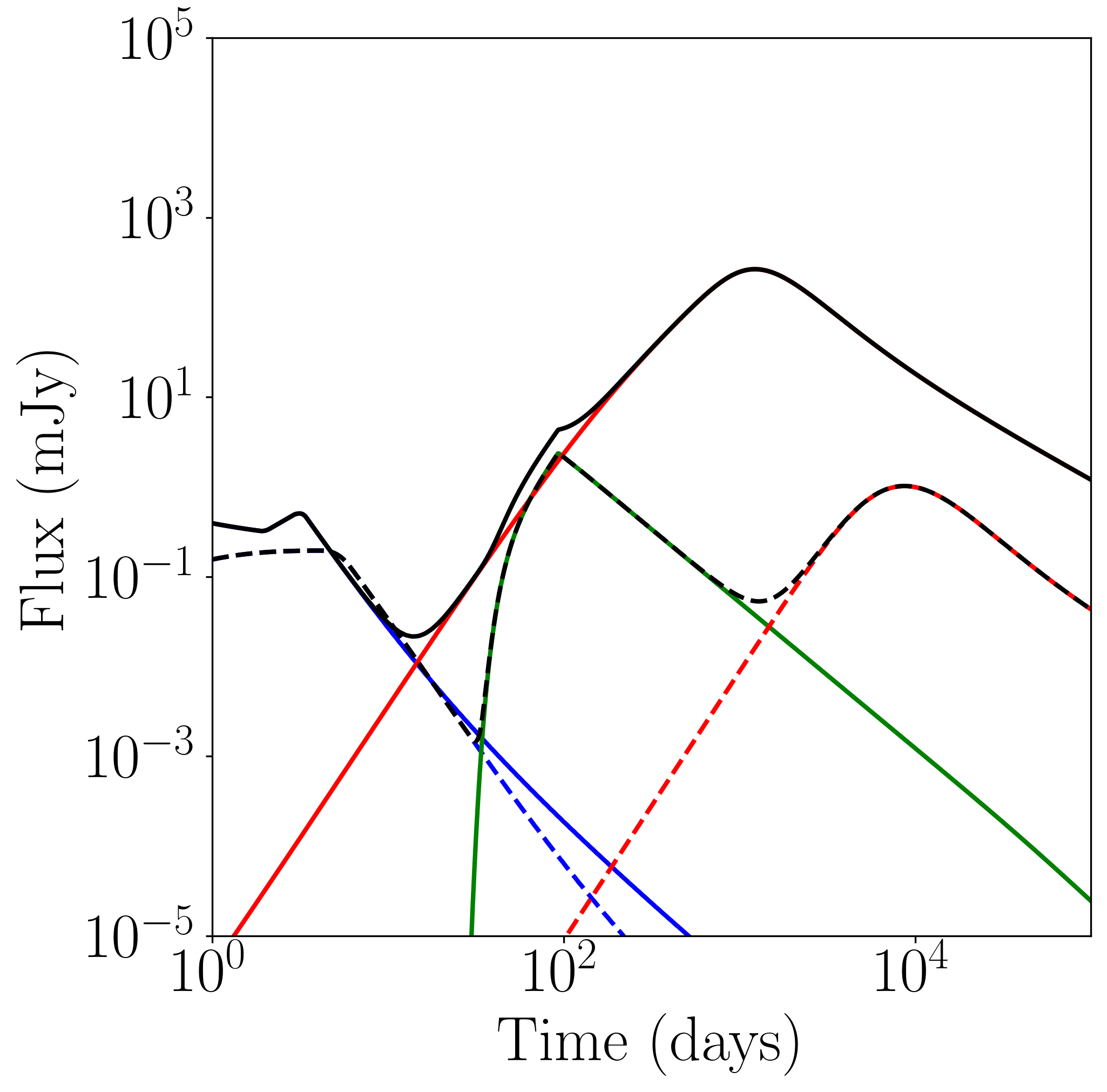}&
\includegraphics[width=1.08\linewidth]{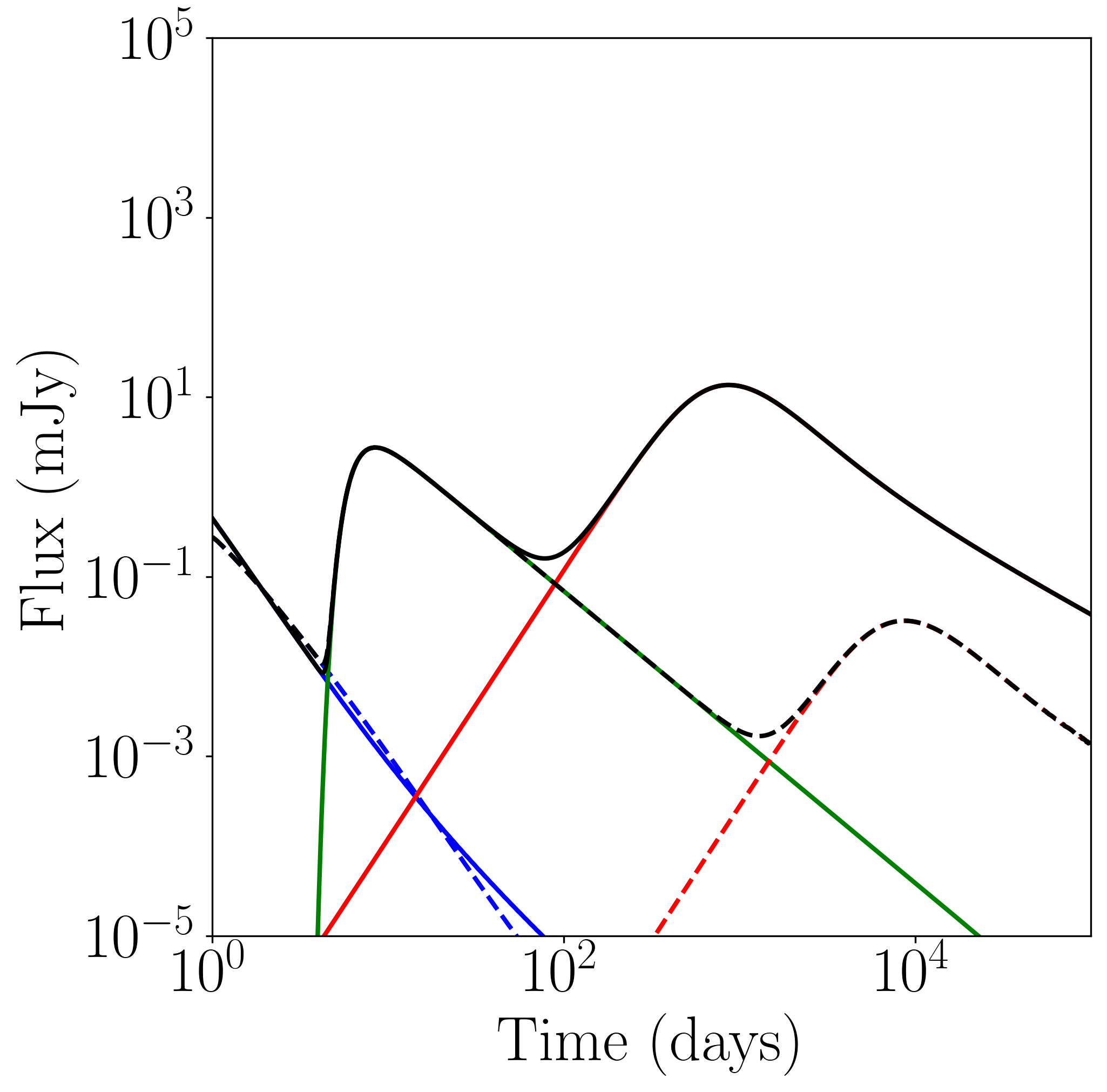}&
\includegraphics[width=1.08\linewidth]{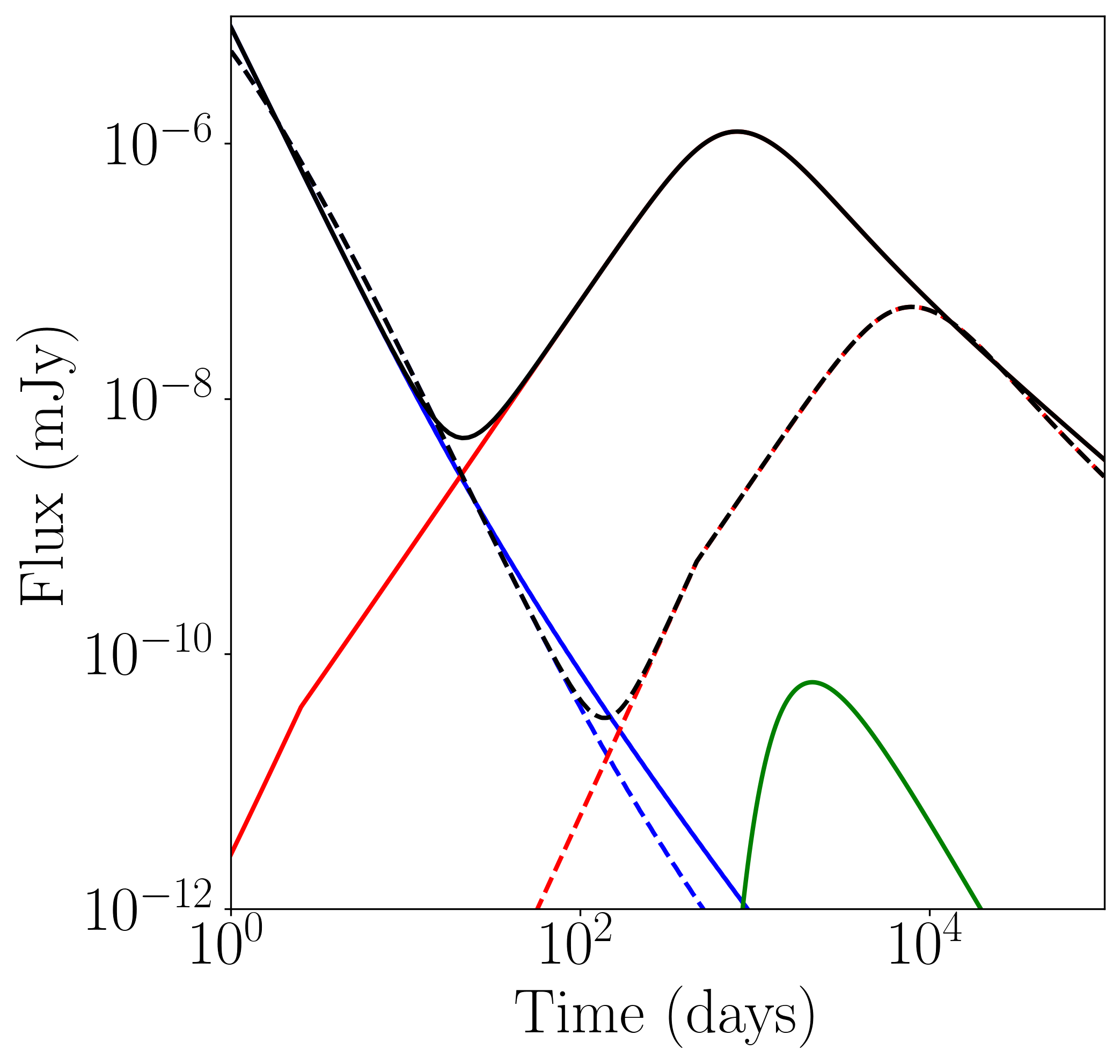}&
\includegraphics[width=1.08\linewidth]{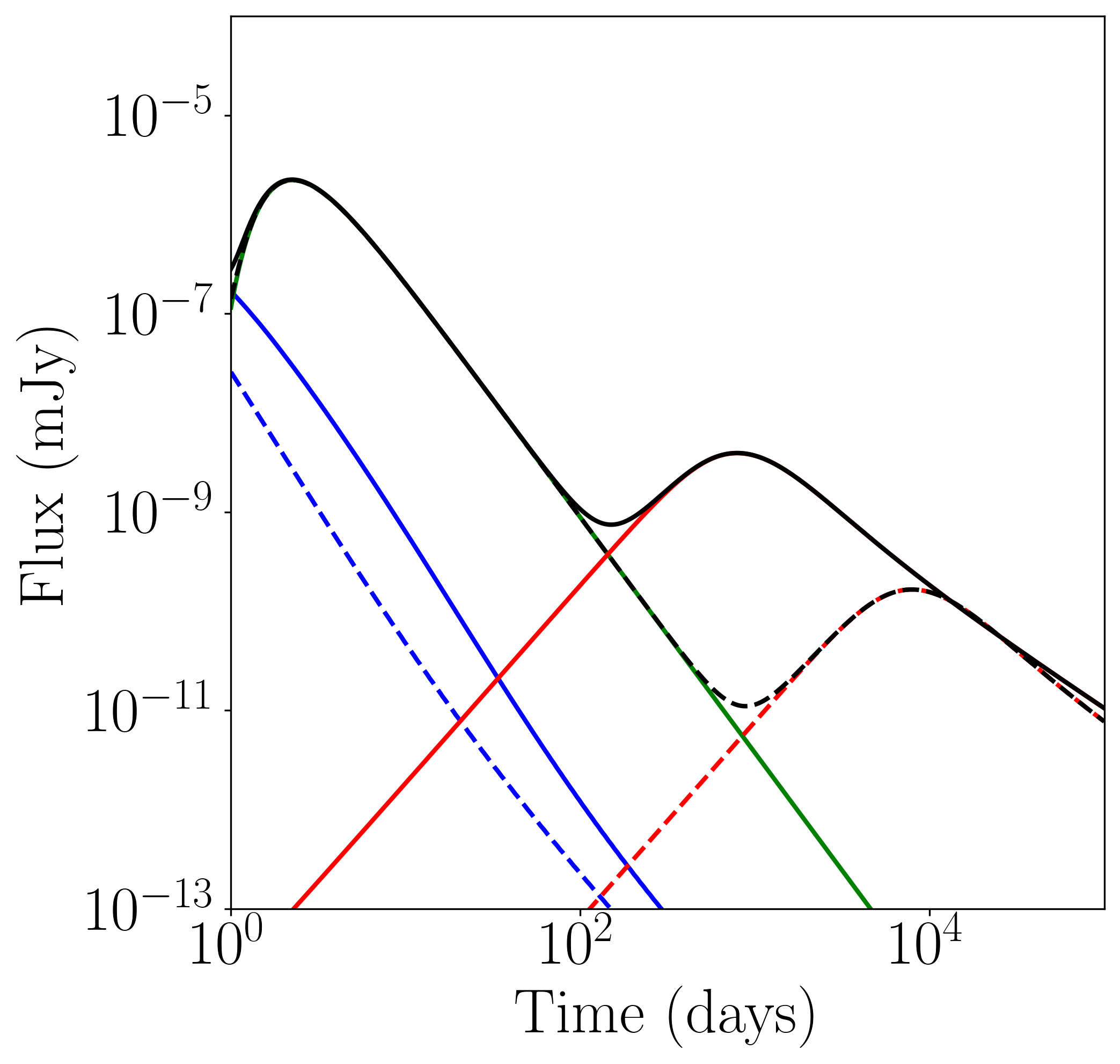}\\[-1.05ex]
\textbf{KN/SGRB (off-axis)}&
\includegraphics[width=1.08\linewidth]{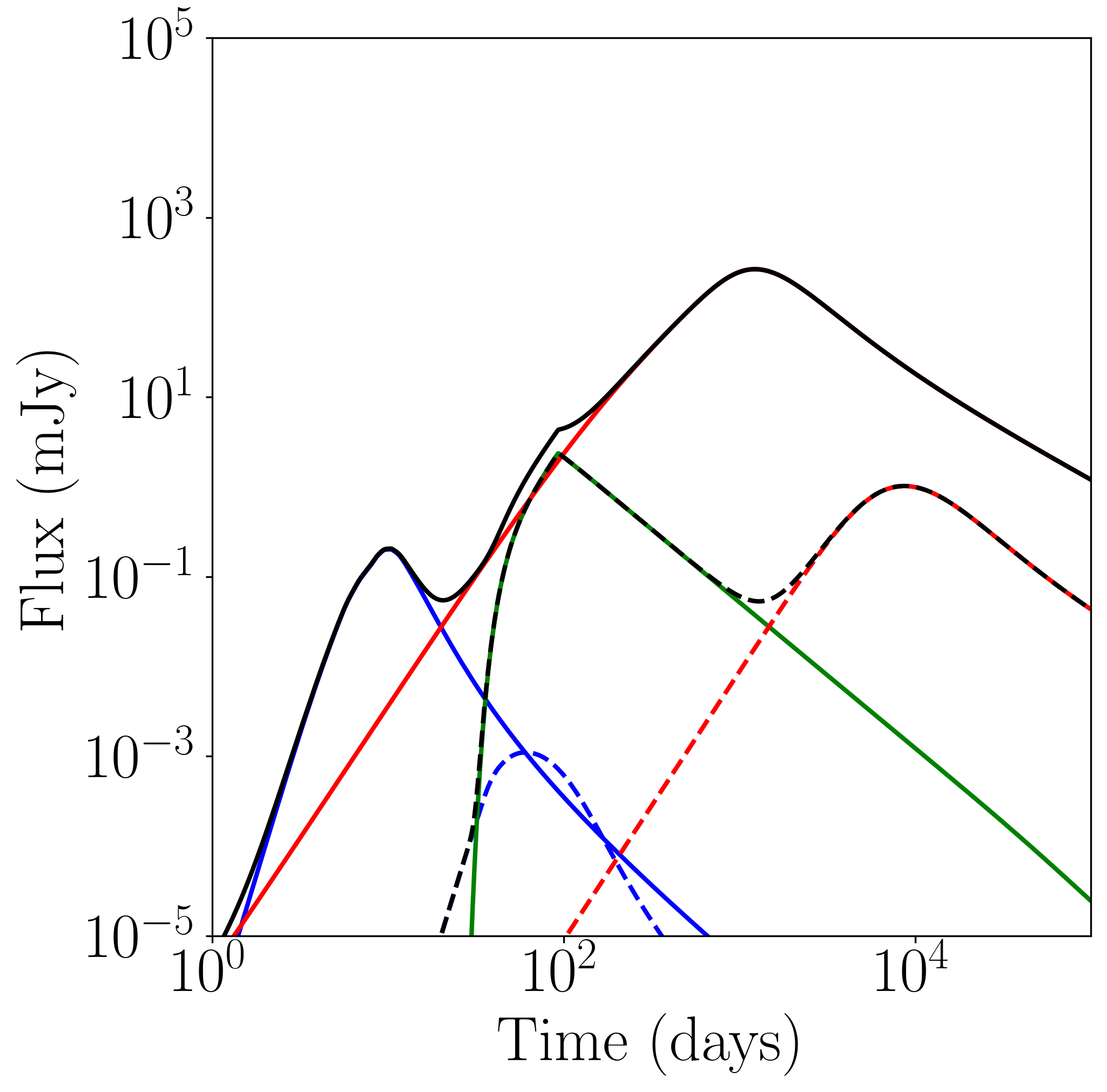}&
\includegraphics[width=1.08\linewidth]{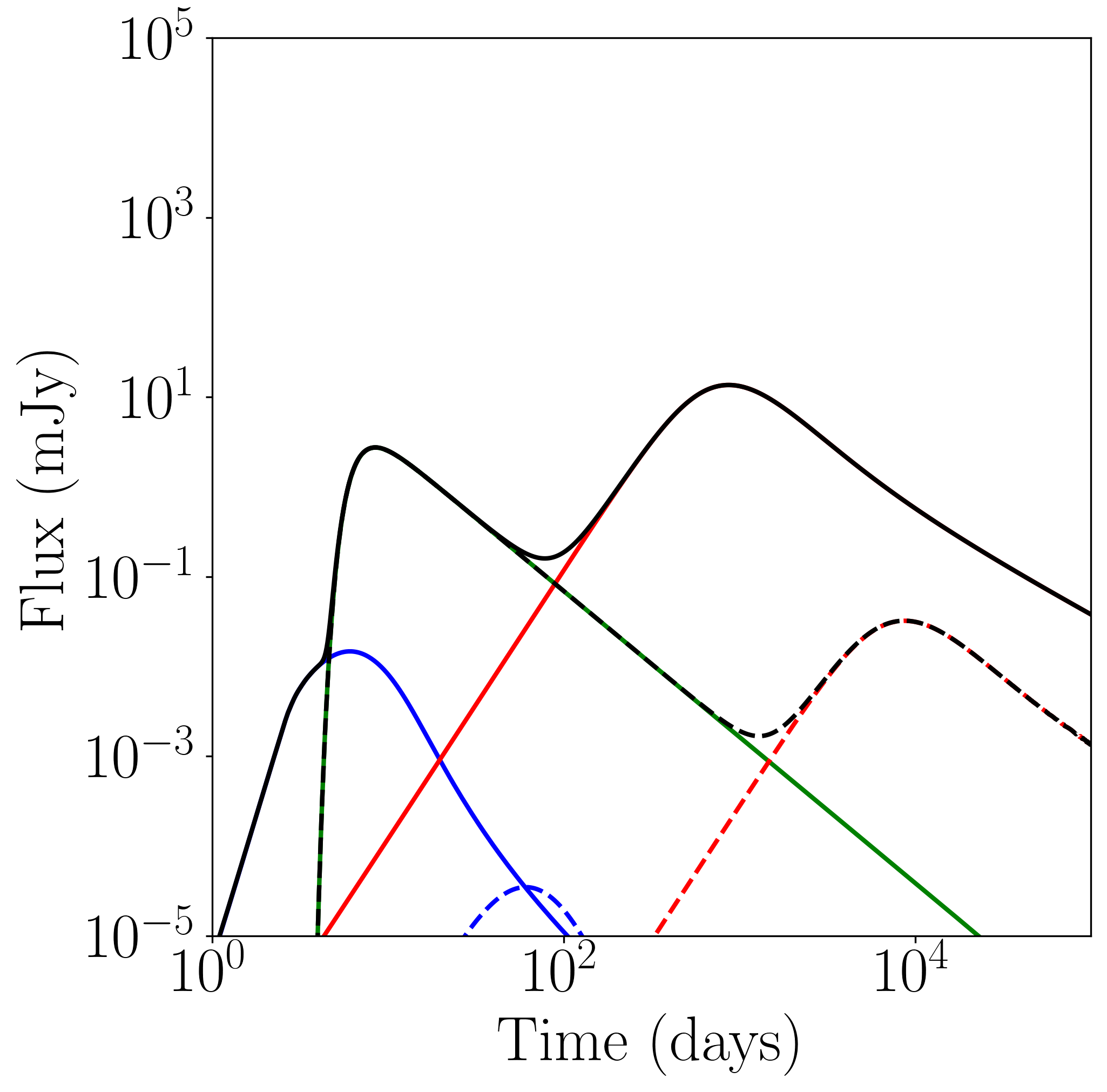}&
\includegraphics[width=1.08\linewidth]{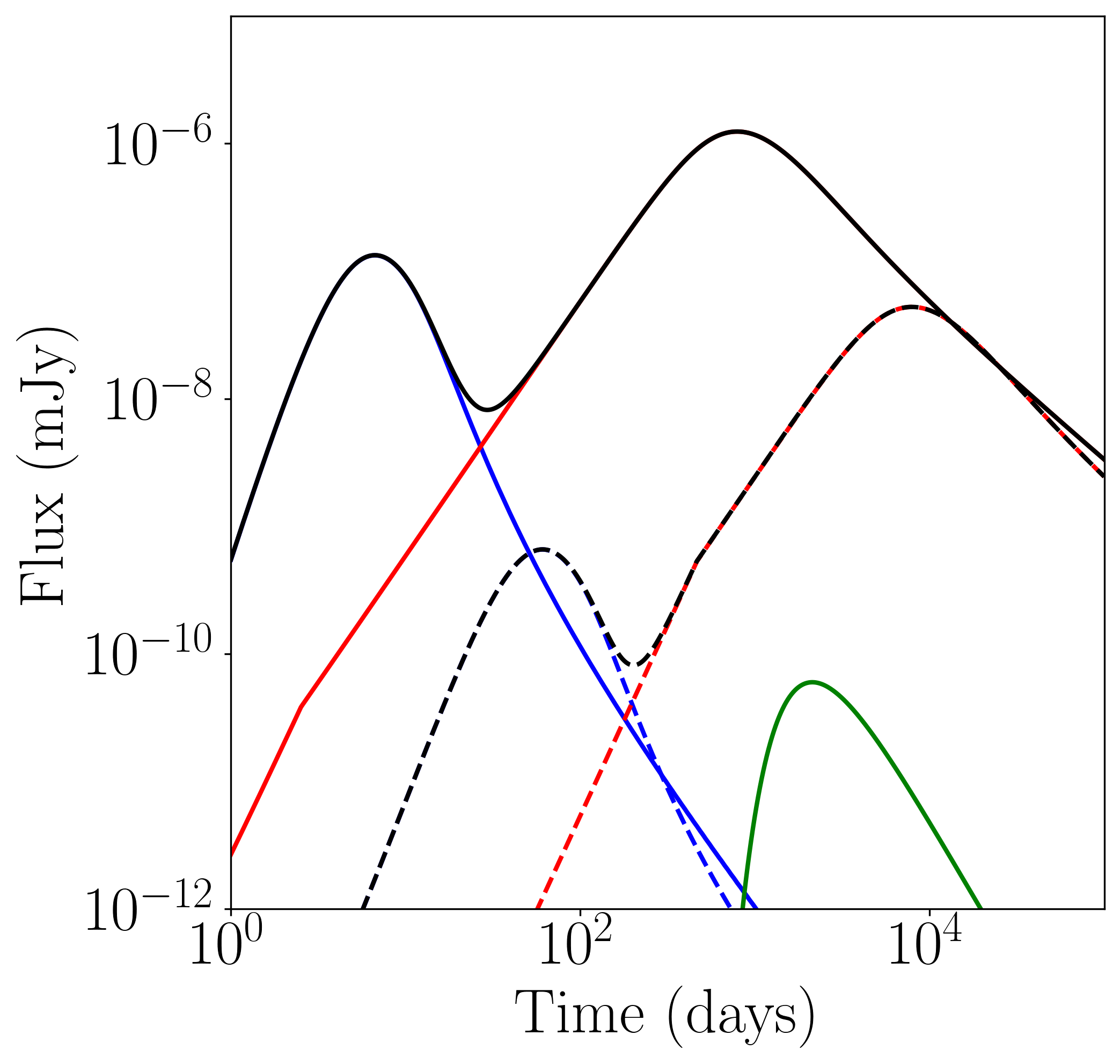}&
\includegraphics[width=1.08\linewidth]{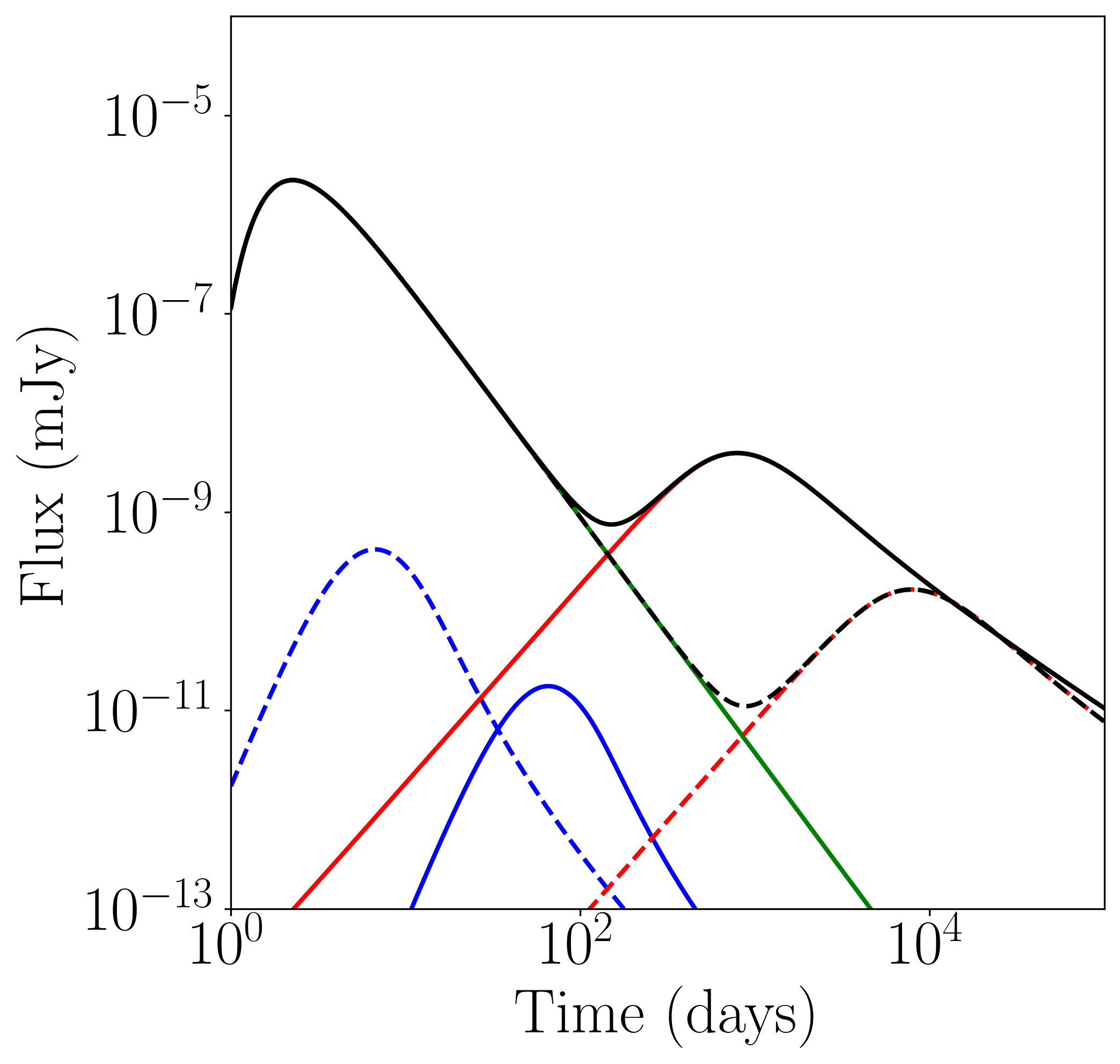}\\[-1.05ex]
\end{tabular}}
\caption{Non-thermal light curves for our fiducial magnetar-driven supernovae and kilonovae, with on- and off-axis GRBs, in radio and X-ray.  Each panel shows the GRB afterglow (blue), PWN (green), ejecta afterglow (red), and total emission (black).  The solid and dashed lines indicate the density of the ambient medium $n_{\rm CSM}$ to be 1 and 10$^{-3}$ cm$^{-3}$ respectively.  The off-axis afterglow is taken from an observer angle of 32\degree.
The transient is assumed to be at 100 Mpc ($z$ = 0.024).} %
\label{fig:examples}
\end{figure*}

Here, we show the fiducial non-thermal light curves of SNe/LGRBs and KN/SGRBs at 1 GHz, 100 GHz, 1 keV, and 100 keV.  The models are generated with \texttt{Redback} \citep{Sarin_redback} using the \texttt{tophat\_redback} \citep{Lamb2018_late}, \texttt{PWN} (This work, Appendix \ref{sec:pwn}), and \texttt{kilonova\_afterglow\_redback} \citep{Margalit2020, Sarin2022} models for the GRB afterglow, PWN, and ejecta afterglow, respectively.  These models for a distance of 100 Mpc are shown in Figure \ref{fig:examples}.

The GRB afterglow model is calculated using a jet opening angle of $\theta_{\rm jet}$ = 0.1 radians, spectral index $p$ = 2.5, initial Lorentz factor $\gamma_0$ = 100, $\epsilon_e$ = 0.1, $\epsilon_B$ = 0.01, and a jet energy of $E_{\rm jet}$ = 10$^{52}$ erg for the LGRB and 10$^{50}$ erg for the SGRB, which are typical parameters inferred from LGRB and SGRB afterglows \citep{Fong2015, Wang2015lgrb, Atteia2017}.  The models are calculated with two values of $n_{\rm CSM}$: 1 cm$^{-3}$ and 10$^{-3}$ cm$^{-3}$, representing the high and low ends of typical observed GRBs \citep{Fong2015, Wang2015lgrb}.  The models are also calculated with two viewing angles, 0\degree \,and 32\degree, which are referred to as on- and off-axis respectively.  The off-axis angle was chosen as 32\degree \, because it is the peak of the inclination distribution for gravitational wave detected sources, and therefore the most likely viewing angle for GW counterparts \citep{Schutz2011, Lamb2017}.

The PWN model is calculated using a magnetar braking index\footnote{The braking index $n$ parameterizes the time dependence of the magnetar spin-down (see Equation \ref{eqn:llasky}).   The value $n = 3$ corresponds to vacuum dipole spin down.} $n = 3$, $\epsilon_B = 10^{-2}$, and electron injection Lorentz factor $\gamma_b = 10^{-5}$, which are typical of Galactic PWNe such as the Crab \citep{Tanaka2010, Tanaka2013} and similar to that used in previous studies of SLSN radio emission \citep[e.g.][]{Omand2018, Law2019, Eftekhari2021}.  The supernova scenario is calculated with initial spin-down luminosity $L_0 = 10^{48}$ erg s$^{-1}$, spin-down timescale $t_{\rm SD} = 10^4$ s, and ejecta mass $M_{\rm ej} = 5 M_\odot$ while the kilonova scenario is calculated with $L_0 = 10^{50}$ erg s$^{-1}$, $t_{\rm SD} = 10^2$ s, and $M_{\rm ej} = 0.05 M_\odot$.  The parameters chosen for the supernova scenario give parameters consistent with previous GRB-SN observations and models of SNe Ic-BL \citep{Taddia2019, Suzuki2021, Omand2024, Srinivasaragavan2024}, while the parameters chosen for the kilonova scenario are motivated by previous modeling efforts and numerical simulations \citep[e.g.][]{Yu2013, Siegel2017, Murase2018, Margalit2019}.  The kilonova is also calculated with a different ejecta composition which is broadly consistent which r-process nucleosynthesis and kilonova spectral simulations \citep{Foucart2016, Roberts2017, Vlasov2017, Hotokezaka2021, Pognan2023, Pognan2025}, with $Y_{\rm e} \sim 0.4$, $\bar{Z} \sim 40$, and $Y_{\rm fe} \sim 0.02$.

The parameters for the ejecta afterglow are mostly shared with the other models, since $n_{\rm CSM}$ is already set in the GRB afterglow model, $M_{\rm ej}$ is set by the PWN model, and $E_{\rm ej}$ is calculated in the PWN model, and set by $L_0$, $t_{\rm SD}$, and $M_{\rm ej}$ (See \citet{Sarin2022} and \citet{Omand2024} for further details).  The partition parameters are set as $\epsilon_e = \epsilon_B = 0.1$, and the spectral index is taken as $p = 2.5$ \citep{Frail2000, Frail2005}.

In radio bands, the PWN is always detectable for the fiducial magnetar-driven supernova.  The viewing angle of the GRB does not impact the timescale of the detectability of the PWN due to the large difference in peak timescales.  The PWN peak timescales of the model are $\sim$ a few hundred days at 100 GHz and $\sim$ a decade at 1 GHz, in agreement with what we derived in Section \ref{sec:timescales}.  The detectability window of the PWN depends on the ambient density, since both afterglow components decrease in luminosity in more rarefied environments, especially the ejecta afterglow.  The GRB afterglow is always the dominant component at early times, and ejecta afterglow at late times.  The PWN is the dominant component after $\sim$ 6 years at 1 GHz and after $\sim$ 1 year at 100 GHz, and it is the dominant component in both bands until $\sim$ 30 years in the high density medium and until $>$ 300 years in the low density medium. 

The radio emission from the PWN in the fiducial magnetar-driven kilonova will be detectable in most situations, although in dense environments, the ejecta afterglow will have comparable luminosity at 1 GHz to the PWN at the PWN peak timescale.  The GRB afterglow is the dominant component at early times if the GRB is on-axis or in a dense ambient medium, but will be subdominant to the PWN in a diffuse medium for an off-axis observer.  The peak timescale is set by synchrotron self-absorption at 1 GHz and free-free absorption at 100 GHz.  The detectability window at 1 GHz in a high-density medium is $\sim$ 40 -- 150 days, and for a low-density medium the window is $\sim$ 30 -- 1500 days, while at 100 GHz the window for a high-density medium is 5 -- 90 days and for a low density medium is 5 -- 1500 days.

In X-rays, there are some regions of the parameter space where the PWN does not dominate the light curve on any timescale in a magnetar-driven supernova, due to the similar peak timescale to the ejecta afterglow.  This is the case at 1 keV in a dense medium, and even in a low-density medium, the PWN will not be the dominant component until several decades post-explosion.  At 100 keV, the PWN will be the dominant component from $\sim$ 50 days until $>$ 30 years, and will be preceded by the GRB in each case except for an off-axis afterglow in a high-density medium, which rises on the same timescale as the PWN but at a lower luminosity.

At 1 keV, the PWN in the fiducial magnetar-driven kilonova will never dominate over the kilonova afterglow.  At 100 keV it dominates from $<$ 1 day until $\sim$ 150 days in a high-density medium and $\sim$ 800 days in a low-density medium.

\subsection{Inference on a Simulated Light Curve} \label{sec:inf}

To determine whether a pulsar wind nebula can be distinguished from an afterglow in data, we simulate and fit an observed radio light curve generated for the fiducial kilonova/SGRB using the {\sc{Redback}} simulation workflow at 1 and 100 GHz for up to 100 days post explosion.  The light curves are observed ten times over the timespan at each frequency.  The distance to the transient is 100 Mpc, giving a redshift of $z$ = 0.024; the neutron star merger that causes this transient would thus also be detectable with current gravitational wave detectors \citep{Abbott2020}.

Inference is performed using {\sc{Redback}}~\citep{Sarin_redback} with the {\sc{pymultinest}} sampler \citep{Buchner2014} implemented in {\sc{Bilby}} \citep{Ashton2019}. 
We sample in flux density with a Gaussian likelihood.
To gauge whether the pulsar wind nebula is needed to explain the observed emission, we implement a Slab-Spike prior \citep{Malsiner-Walli2018} on the initial pulsar wind nebula luminosity $L_0$, which imposes a dirac-delta function onto an already existing prior.  We give the spike 10\% of the probability of the prior and place it at the lowest value of $L_0$ within the prior.  Having a large probability of a low value of $L_0$ allows us to test if the emission can be reproduced with the PWN effectively turned off.  A list of the parameters varied in the inference, as well as the median and 1$\sigma$ values of the posterior, is found in Table \ref{tbl:priors}.  Other parameter values are kept constant at their fiducial values (see Section \ref{sec:typical}).

\begin{table*}
\centering
\begin{tabular}{cccccc}
   Parameter & Definition & Units & Injected Value & Prior & 1D Posterior Values \\ \hline
   $\theta_{\rm v}$ & Viewing Angle & Radians & 0.1 & Sine[0,$\pi/2$] & 0.34$^{+0.10}_{-0.14}$ \\
   $E_{\rm jet}$ & Jet Energy (Isotropic Equivalent) & erg & $10^{50}$ & L[$10^{46}$, $10^{53}$] & $L$(49.87$^{+0.65}_{-0.55}$) \\
   $\epsilon_{\rm e, GRB}$ & Lepton Energy Parameter (GRB) & & 0.1 & L[$10^{-5}$, 1] & $L$(-1.07$^{+0.44}_{-1.30}$) \\
   $\epsilon_{\rm B, GRB}$ & Magnetization Parameter (GRB) & & 0.01 & L[$10^{-5}$, 1] & $L$(-1.38$^{+0.95}_{-1.68}$) \\
   $n_{\rm CSM}$ & CSM number density & cm$^{-3}$ & 10$^{-2}$ & L[10$^{-5}$, 10$^{2}$] & $L$(0.74$^{+0.75}_{-1.15}$) \\
   $L_0$ & Initial Magnetar Spin-Down Luminosity & erg s$^{-1}$ & 10$^{50}$ & S[$10^{40}$, L[$10^{40}$, $10^{51}$]] & $L$(45.46$^{+1.15}_{-1.01}$) \\
   $t_{\rm SD}$ & Spin-Down Time & s & 10$^2$ & L[$10$, $10^{8}$] & $L$(4.35$^{+0.69}_{-0.90}$) \\
   $M_{\rm ej}$ & Ejecta Mass & $M_{\odot}$ & 0.05 & L[0.01, 100] & 0.02$^{+0.00}_{-0.00}$ \\
   $n$ & Magnetar Braking Index & & 3 & U[1.5, 10] & 2.80$^{+0.26}_{-0.25}$ \\
   $\epsilon_{\rm B,PWN}$ & Magnetization Parameter (PWN) & & 10$^{-2}$ & L[10$^{-7}$, 1] & $L$(-1.29$^{+0.60}_{-0.56}$) \\
   $\gamma_b$ & Electron Injection Lorentz Factor & & 10$^5$ & L[10$^2$, 10$^8$] & $L$(4.77$^{+1.73}_{-1.55}$) \\
   $\epsilon_{\rm e, ejecta}$ & Lepton Energy Parameter (Ejecta) & & 0.1 & L[$10^{-5}$, 1] & $L$(-3.41$^{+1.47}_{-1.05}$) \\
   $\epsilon_{\rm B, ejecta}$ & Magnetization Parameter (Ejecta) & & 0.1 & L[$10^{-5}$, 1] & $L$(-3.21$^{+1.67}_{-1.18}$) \\ \hline
\end{tabular}
\caption{The parameters and priors used in this study. Priors are either uniform (U) log-uniform (L), sine (Sine), or Slab-Spike (S).  The values shown for the posterior are the median and 1$\sigma$ uncertainties.  Posterior values denoted with $L$ are given in log-space.  The full posterior is shown in Figure \ref{fig:sim_corner}.}
\label{tbl:priors}
\end{table*}

\begin{figure}
    \centering
    \includegraphics[width=\linewidth]{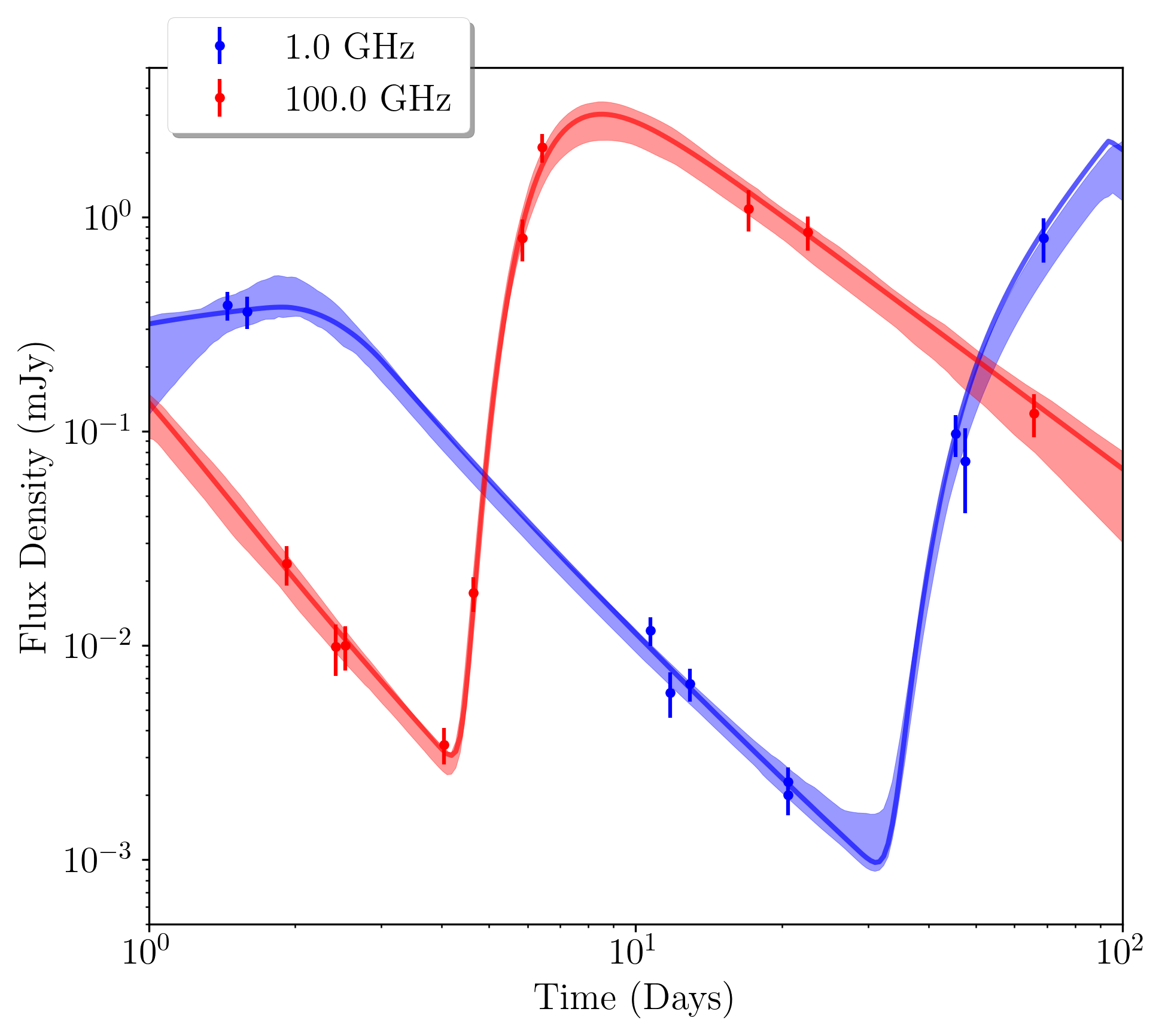}
    \caption{The fitted radio light curve for the simulated kilonova/SGRB for both 1 and 100 GHz.  The solid line shows the model with the highest likelihood while the shaded region shows the 90$\%$ confidence interval.}
    \label{fig:sim_fit}
\end{figure}

The fitted data is shown in Figure \ref{fig:sim_fit} and the posterior is shown in Figure \ref{fig:sim_corner}.  The posteriors for the afterglow parameters $E_{\rm jet}$, $\epsilon_{\rm e, GRB}$, $\epsilon_{\rm B, GRB}$ and as well as $n$, $\epsilon_{\rm B, PWN}$, and $\gamma_b$, are all found to be within around 1$\sigma$ of the injected parameters.  $\theta_{\rm obs}$ and $n_{\rm CSM}$ are more than 1$\sigma$ away, but are correlated, while $\epsilon_{\rm e, ejecta}$ and $\epsilon_{\rm B, ejecta}$ are not well constrained due to the signal lacking an ejecta afterglow component.  The one-dimensional posteriors for $L_0$ and $t_{\rm SD}$ underestimate and overestimate the injected values respectively, although the degeneracy between them is clear in the two-dimensional posterior.  The ejecta mass is also significantly underestimated.  These three parameters all show strong correlation and have wide priors with the injected value close to the edge of the prior.  The escape time (Equation \ref{eqn:tesc_ff}) shows a dependence on both ejecta mass and velocity, so the lower inferred ejecta mass implies that the ejecta in the inferred model must have lower velocity than in the injected model, which is consistent with a lower $L_0$ and higher $t_{\rm SD}$.  This example shows the sensitivity of the inferred parameters to the intrinsic noise of the measurement.  Other estimates of $M_{\rm ej}$ or $v_{\rm ej}$ in a real system, either from gravitational wave data \citep{Abbott2019prop}, kilonova light curve modeling \citep{Yu2013, Metzger2019, Sarin2022}, or kilonova spectroscopy \citep{Hotokezaka2021, Pognan2023}, would help get a more accurate estimate of the PWN and ejecta parameters.

The posterior for $L_0$ shows no probability at 10$^{40}$ erg, showing that it is unlikely that this emission can be reproduced without a strong PWN.  The main features that can not be reproduced by an ejecta afterglow are the frequency-dependent peak timescales and the fast rise from only GRB afterglow emission to PWN peak.  The ejecta afterglow emission timescale can be frequency-dependent if either the self-absorption frequency or synchrotron frequency of the minimum Lorentz factor is higher than the observed frequency, which can happen at low frequencies in atypical cases.  This would cause a dependence of either $t_{\rm peak} \propto \nu_{\rm obs}^{-1/3}$ if $\nu_m$ is higher or $t_{\rm peak} \propto \nu_{\rm obs}^{-(4+p)/(3p-2)}$ if $\nu_{\rm ssa}$ is higher \citep{Nakar2011}, which could mimic the $t_{\rm peak} \propto \nu_{\rm obs}^{-0.42}$ expected for a PWN.  However, the rising flux for an ejecta afterglow is a power law which becomes slower ($F_\nu \approxprop t^{1.5}$ as opposed to $F_\nu \approxprop t^{3}$) under the conditions where $t_{\rm peak}$ becomes frequency-dependent \citep{Nakar2011}.  The ejecta afterglow rising flux also contrasts with the PWN rising flux, which is an exponential rise due to being caused by optical depth effects.  While this can be used to distinguish between the two scenarios, one needs multiple extremely high cadence observations during the light curve rise to distinguish them via the rise index alone.  Another method of confirming the presence of a PWN is to continue observing on $\sim$ year timescales when the ejecta afterglow component dominates, as the presence of the third peak would provide further evidence for this scenario.  However, this is only viable for SGRBs, as the ejecta afterglow is not expected to dominate the emission for LGRBs until decades after the burst.

\section{Discussion} \label{sec:disc}

Although the PWN can be the dominant emission component at certain times and frequencies, these PWNe are still not bright enough to be seen at large redshifts.  The detection limit for NuSTAR in hard x-rays is roughly 1 nJy \citep{Vurm2021}, which gives a detection horizon of only $\sim$ 30 Mpc for our fiducial SN/LGRB PWNe and 3 Mpc for our fiducial KN/SGRB PWNe.  The Karl G. Jansky Very Large Array (VLA) and Atacama Large Millimeter/submillimeter Array (ALMA) have 3$\sigma$ detection limits of roughly 15 and 50 $\mu$Jy for 1 and 100 GHz respectively \citep{Omand2018, Eftekhari2021}.  Based on Figure \ref{fig:examples}, the detection horizon for our fiducial SN/LGRB PWNe is $\sim$ 250 Mpc ($z \sim 0.06$) at 1 GHz and 200 Mpc ($z \sim 0.05$) at 100 GHz, while for our fiducial KN/SGRB PWN the detection horizon is $\sim$ 1.5 Gpc ($z \sim 0.3$) at 1 GHz and 1.0 Gpc ($z \sim 0.2$) at 100 GHz.  The horizon for SNe/LGRBs is lower than almost all Ic-BL SNe \citep{Taddia2019, Srinivasaragavan2024} and LGRBs \citep{Horvath2022}, however, the horizon for KNe/SGRBs contains roughly 10$\%$ of the SGRB distribution \citep{Ghirlanda2016}.  \citet{Schroeder2020} observed several $z < 0.5$ SGRBs at 6 GHz around $2-14$ years post-burst and did not detect any emission, placing constraints on both the PWN and ejecta afterglow emission.

Next-generation radio telescopes, such as DSA-2000 and ngVLA will have 3$\sigma$ detection limits of $\sim$ 1 $\mu$Jy in the 1 -- 100 GHz bands \citep{Hallinan2019, McKinnon2019, DiFrancesco2019}, giving horizons of $\sim$ 1 -- 1.5 Gpc ($z \sim 0.2-0.3$) for our fiducial SN/LGRB PWNe and $\sim$ 4 -- 10 Gpc ($z \sim 0.7-1.5$) for our fiducial KN/SGRB PWNe.  Most classified Ic-BL SNe are detected below the SNe/LGRB horizon \citep{Taddia2019, Srinivasaragavan2024}, although the majority of LGRBs are at higher redshifts \citep{Lan2021}.  The horizon for SGRBs comprises around 50\% of the population at 1 GHz and almost the entire population at 100 GHz \citep{Ghirlanda2016}.

We precluded modeling of the optical/UV signal in Section \ref{sec:typical} due to possible contamination from the associated SN/KN.  The thermal SN/KN component can be modeled along with the non-thermal components (see e.g. \citet{Wallace2024} for joint afterglow/kilonova modeling), but we did not want to present four-component models for simplicity.  The absorption processes for optical/UV emission are primarily bound-bound transitions that thermalize the energy in the ejecta, and the escape timescale is thus given by the nebular timescale for the ejecta.  For typical GRB-SNe, this timescale is around 1 -- 2 months, while for magnetar-driven KNe, this timescale is around 1 -- 5 days.  Models of these transients predict strong optical emission at these times \citep{Sarin2022, Omand2024}, although near-UV emission can decline sharply as the transient cools, leaving a window for the non-thermal emission to be detectable at these wavelengths.

Given the need for well-timed, high-cadence, multi-band observations to confirm this scenario, it's important to note the early signals of magnetar-driven GRBs.  This is especially important to coordinate facilities like ALMA, which can take as long as two weeks to respond to a target-of-opportunity trigger\footnote{\url{https://almascience.eso.org/observing/too-activation}}, which is comparable to the escape timescale at 100 GHz in the fiducial KN/SGRB.  With the horizons calculated above, it is unlikely we will detect PWN emission for all but the closest LGRBs or for SGRBs above $z \approx 0.3$ with current instruments, so it is best to focus on low-redshift SGRBs.  Given the potential overlap of the PWN and ejecta afterglow components at lower radio frequencies for SGRBs, targeting the afterglows with high frequency radio observations has a higher chance of a clear discovery, which can be followed up with lower frequency radio observations to further characterize the PWN and ejecta afterglow components. Early signals typically associated with the presence of a central magnetar are x-ray plateaus \citep{Rowlinson2013, Gompertz2014, Stratta2018, Sarin2020quark} and extended emission \citep{Metzger2008, Bucciantini2012, Gompertz2013, Gompertz2014, Gibson2017, Sarin2020radloss}.  For closer objects, a gravitational wave detection of a binary neutron star merger coincident with the GRB showing a low chirp mass and an anomalously bright kilonova \citep{Yu2013, Metzger2019, Sarin2022, Ai2024}, would mark it as a strong candidate for being magnetar-powered.  Binary neutron star mergers are also expected to be neutrino sources on a short timescale \citep{Kyutoku2018}, but a long-lived magnetar remnant could emit neutrinos on a longer timescale due to the interaction of the hadrons in the pulsar wind and ejecta \citep{Bednarek2003, DiPalma2017}.

The presence of a magnetar engine can be inferred from the luminosity and timescale of the thermal transient or the detection of non-thermal emission from the PWN, but the magnetar may also have other detectable effects.  \citet{Omand2019} suggested that dust formed in the ejecta of magnetar-driven supernovae can be heated by the PWN and produce an observable signal in the infrared.  While this signal could be present in GRB-SNe, those systems could also have an infrared excess due to r-process nucleosynthesis in a collapsar \citep{Siegel2019, Barnes2022, Anand2024}, so this signal would not be definitive.  In kilonovae with no energy injection into the ejecta, dust is difficult to form due to the low density of the ejecta \citep{Takami2014}; adding energy injection would only make this problem worse due to the slower cooling and faster ejecta velocity, so this is likely not a viable method of detecting magnetars in kilonovae.  Also, kilonovae are intrinsically bright in the infrared, so an infrared excess from dust would be difficult to disentangle from the kilonova emission.  The PWN can also ionize the ejecta, leading to higher ionization lines in the nebular optical/UV spectrum of the transient.  This effect has been examined in stripped-envelope supernovae \citep{Chevalier1992, Dessart2019, Omand2023, Dessart2024}, with recent efforts focused on replicating the spectrum of possible magnetar-driven supernova SN 2012au \citep{Milisavljevic2013, Milisavljevic2018}.  However, these spectral models are still missing key physics to make accurate predictions for these systems (see \citet{Omand2023} for details).  The ejecta velocity of a magnetar-driven kilonova is expected to be $\gtrsim 0.4c$ \citep{Sarin2022}, so most lines would likely be broadened too much to be detectable.  Models of nebular spectra are also extremely uncertain due to a lack of atomic data \citep{Pognan2023, Hotokezaka2023, Banerjee2025, Pognan2025}, so increasing the complexity of those models with a central energy source would likely be unfeasible for the foreseeable future.  A PWN could also induce polarization in the ejecta either by injecting energy asymmetrically \citep{Inserra2016, Saito2020} or by causing hydrodynamic instabilities in the ejecta \citep{Chen2016, Suzuki2017, Suzuki2021}.  Modeling of polarization in similar systems is scarce \citep{Tanaka2017, Bulla2019} and their polarization without a central magnetar is not well understood, so a polarization measurement will not be able to constrain the presence of a magnetar without further modeling.

Modeling the three different emission components and their interactions is extremely difficult, and our simplified treatment of the non-thermal emission components has a few notable caveats.  The PWN and transient ejecta are assumed to be spherical in this model, although in reality this symmetry can be broken in plenty of ways.  Most numerical simulations of kilonovae show large deviations from spherical symmetry \citep{Bauswein2013, Hotokezaka2013, Rosswog2014}, and interaction between the jet and ejecta can cause the geometry and emission of both to be affected \citep{Nativi2021, Nativi2022}.  The PWN can be aspherical due to Kelvin-Helmholtz instabilities, and the nebula can also cause strong Rayleigh-Taylor instabilities within the transient ejecta, which can form ejecta filaments \citep{Davidson1985, Jun1998, Bucciantini2004, Porth2014} and cause the PWN forward shock to break out of the material \citep{Blondin2017, Suzuki2017, Omand2024Crab}.  The rebrightenings could also be caused by more complicated jet geometries and afterglow physics, such as an exotic jet shape \citep{Takahashi2021, Beniamini2022}, counterjet \citep{Granot2003, Li2004, Wang2009, Zhang2009, GhoshDastidar2024}, reverse shock \citep{Kobayashi1999, Kobayashi2000, Uhm2012}, or late energy injection into the shock, causing a refreshed shock \citep{Rees1998, Panaitescu1998, Zhang2002}.  However, the peak timescales for reverse shocks and counterjets are expected to be much shorter \citep{Kobayashi2000, Uhm2012} and longer \citep[e.g.][]{vanEerten2012}, respectively. The GRB and ejecta afterglow models we use both assume the surrounding medium has constant density, which is not true in the case of wind or eruptive mass loss from the progenitor.  The top-hat model also ignores the effects of the magnetar on the jet itself, which could produce an x-ray plateau in the early afterglow \citep{Rowlinson2013, Gompertz2014, Stratta2018}.  The use of a one-zone spherical ejecta afterglow also neglects the effects of the density structure of the ejecta, which can cause over an order of magnitude difference in the observed signal \citep{Rosswog2024}.

Radio and hard X-ray opacities are relatively straightforward to compute, but our treatment of the soft X-ray opacity from photoabsorption (Equation \ref{eqn:kpe}) neglects a lot of complicated physics.  In general, the photoabsorption opacity depends on both the composition and time-dependent ionization state of the ejecta, which depends on the luminosity and spectrum of the ionizing PWN radiation and the recombination rates from the relevant ions.  We include some compositional dependence with $\bar{Z}$, but neglect the changes in opacity due to changes in ionization, which require detailed radiative transfer simulations to compute.  \citet{Kashiyama2016} uses the same treatment and notes that it likely overestimates the photoabsorption opacity.  The PWN can drive an ionization front through the ejecta, which can produce UV/soft X-ray emission if it breaks out; \citet{Metzger2014ibo} and \citet{Metzger2014kn} discuss this case for magnetar-driven SLSNe and KNe, respectively, and outline the conditions where it can happen.  X-ray surveys of SLSNe have shown soft X-ray emission in some cases \citep{Levan2013}, although this emission is rare overall \citep{Margutti2018}.  Nebular spectra of potential magnetar-driven supernovae generally show line emission from species that would be ionized out given a soft X-ray breakout, so this phenomenon must be rare.  Examining Figure \ref{fig:examples} shows that a decrease in the soft X-ray opacity may lead the PWN to dominate the emission on ~1000 day timescales for the SN/LGRB case, but it is still unlikely for the PWN to dominate in the KN/SGRB case unless the ambient density is extremely low due to the relative brightness of the ejecta afterglow.  However, this emission is not predicted to be detectable by something like Chandra unless the transient is extremely close.

The parameters used for the microphysics of our fiducial PWN are consistent with that of Galactic PWNe such as the Crab \citep{Tanaka2010, Tanaka2013}, the lack of observed radio counterparts \citep{Law2019, Eftekhari2021} shows that this assumption may not be correct for most magnetar-driven supernovae.  A couple of alternate microphysical models are the high-magnetization PWN model \citep{Murase2021}, where the synchrotron energy peaks in the MeV range, and the low-magnetization model \citep{Vurm2021}, where the emission is dominated by inverse-Compton emission ($Y_{\rm IC} \gtrsim 10$).  The latter of these models is also motivated by strong non-thermal losses in SLSNe \citep{Vurm2021}.  The radio emission from both of these cases would be significantly lower than the Crab-like case \citep{Murase2021}.

\begin{figure}
    \centering
    \includegraphics[width=\linewidth]{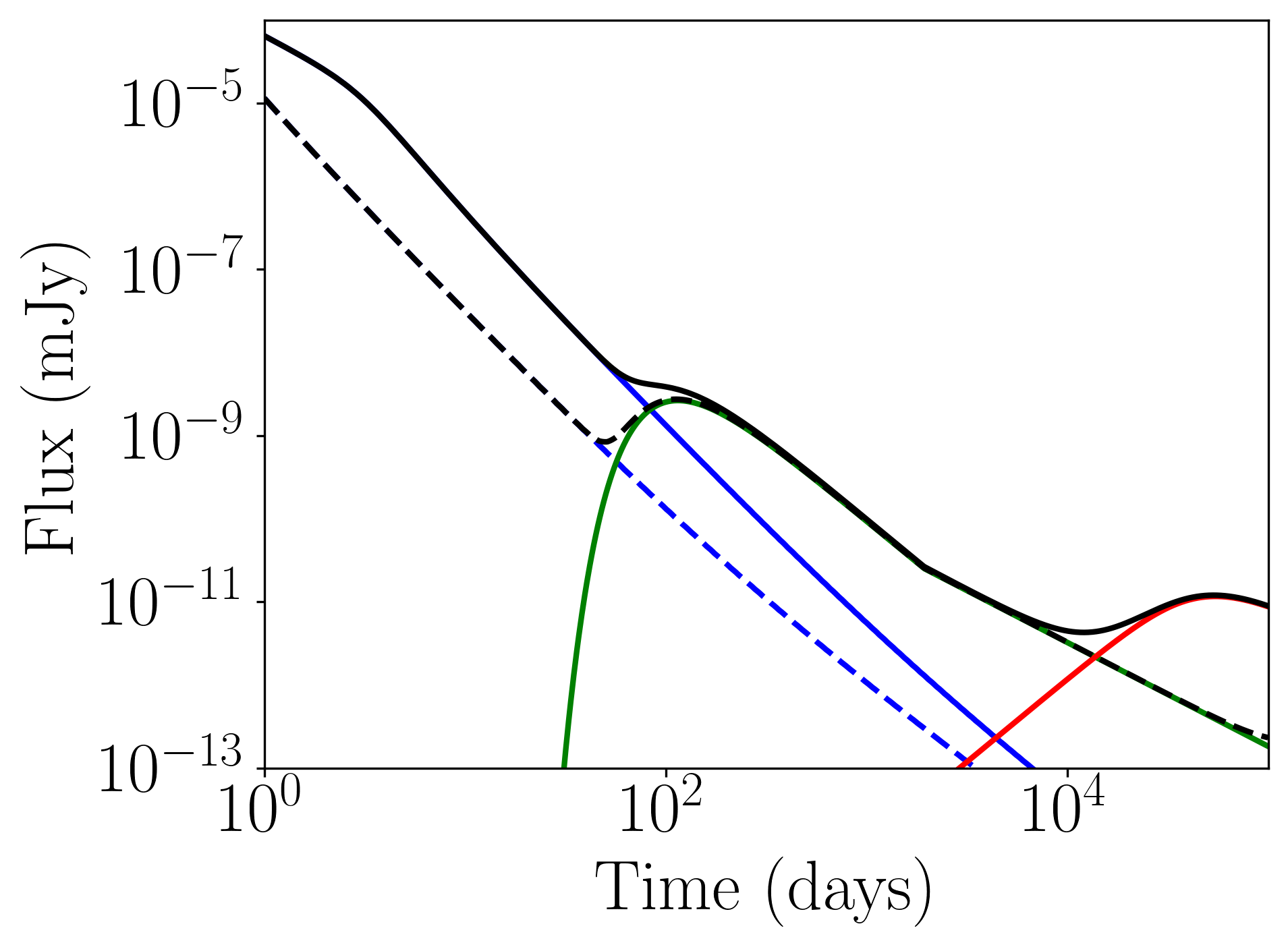}
    \includegraphics[width=\linewidth]{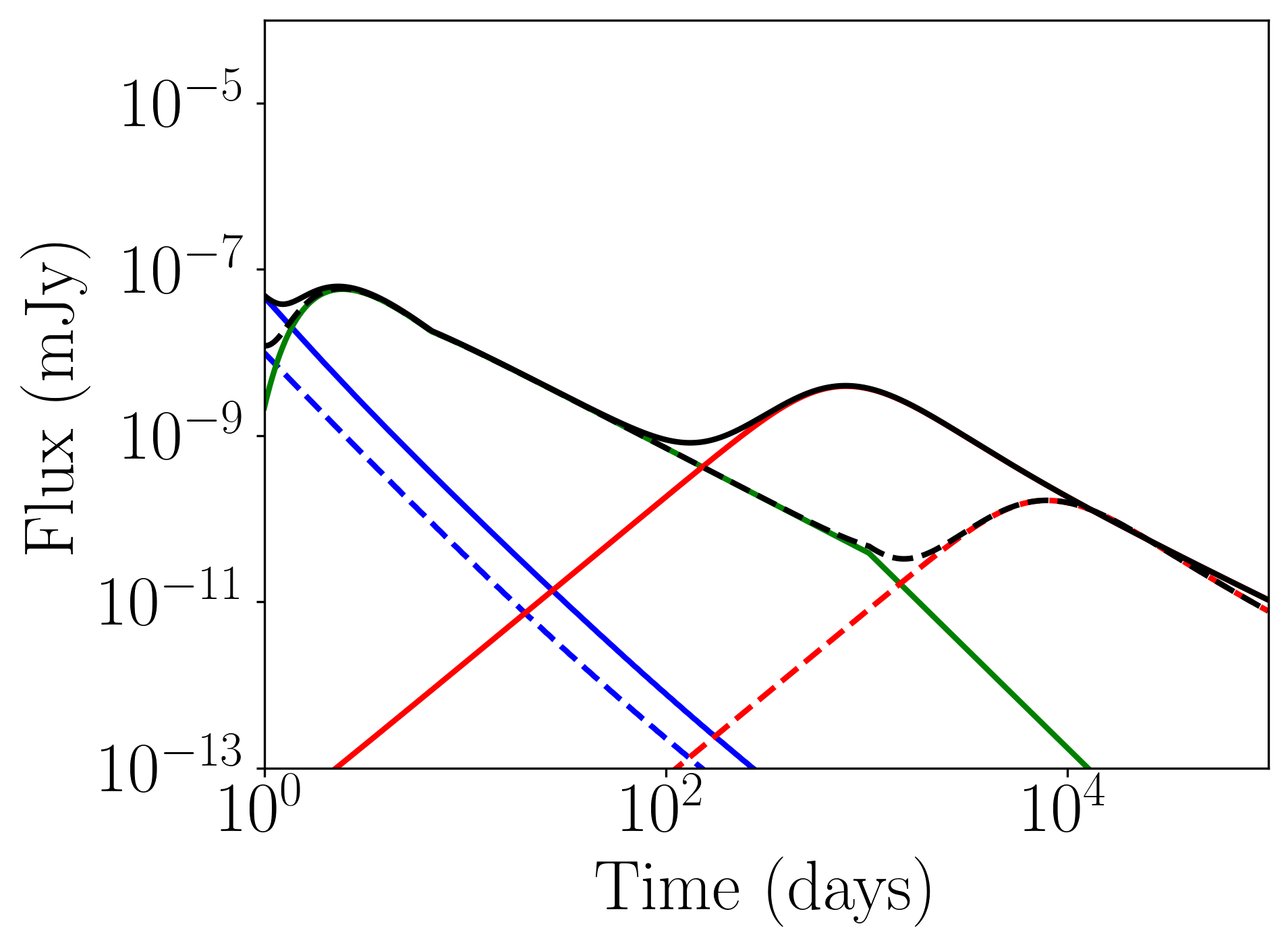}
    \caption{100 keV light curves for magnetar-driven supernovae and kilonovae with an on-axis GRB and a high-magnetization PWN.  Each panel shows the GRB afterglow (blue), PWN (green), ejecta afterglow (red), and total emission (black).  The solid and dashed lines indicate the density of the ambient medium $n_{\rm CSM}$ to be 1 and 10$^{-3}$ cm$^{-3}$ respectively.}
    \label{fig:highmag}
\end{figure}

Because our model does not include inverse Compton emission for simplicity, we cannot treat the low-magnetization model, but we can examine the high-magnetization model further.  Figure \ref{fig:highmag} shows the 100 keV light curves for our fiducial SN/LGRB and KN/SGRB with $\epsilon_{\rm B, PWN} = 0.5$, $\gamma_b = 10^7$, and the low-energy spectral index $q_1 = 1$, as used in \citet{Murase2021}.  The PWN signal at 100 keV for both types of objects is lower than the cases with Crab-like microphysics due to the peak of the spectrum being in the 10 -- 100 MeV range and the steeper spectral index.  This is consistent with \citet{Vurm2021} and \citet{Murase2021}, which show the two models having emission of around the same order of magnitude at 100 keV.  Those studies also show that detecting the emission at $>$ 1 MeV will likely require next generation instruments, such as eASTROGAM and AMEGO.  While it may be possible to have a synchrotron spectrum that peaks at $\sim$ 100 keV around the PWN escape time, such a model would likely need to be fine-tuned and require more physical motivation.

\begin{figure}
    \centering
    \includegraphics[width=\linewidth]{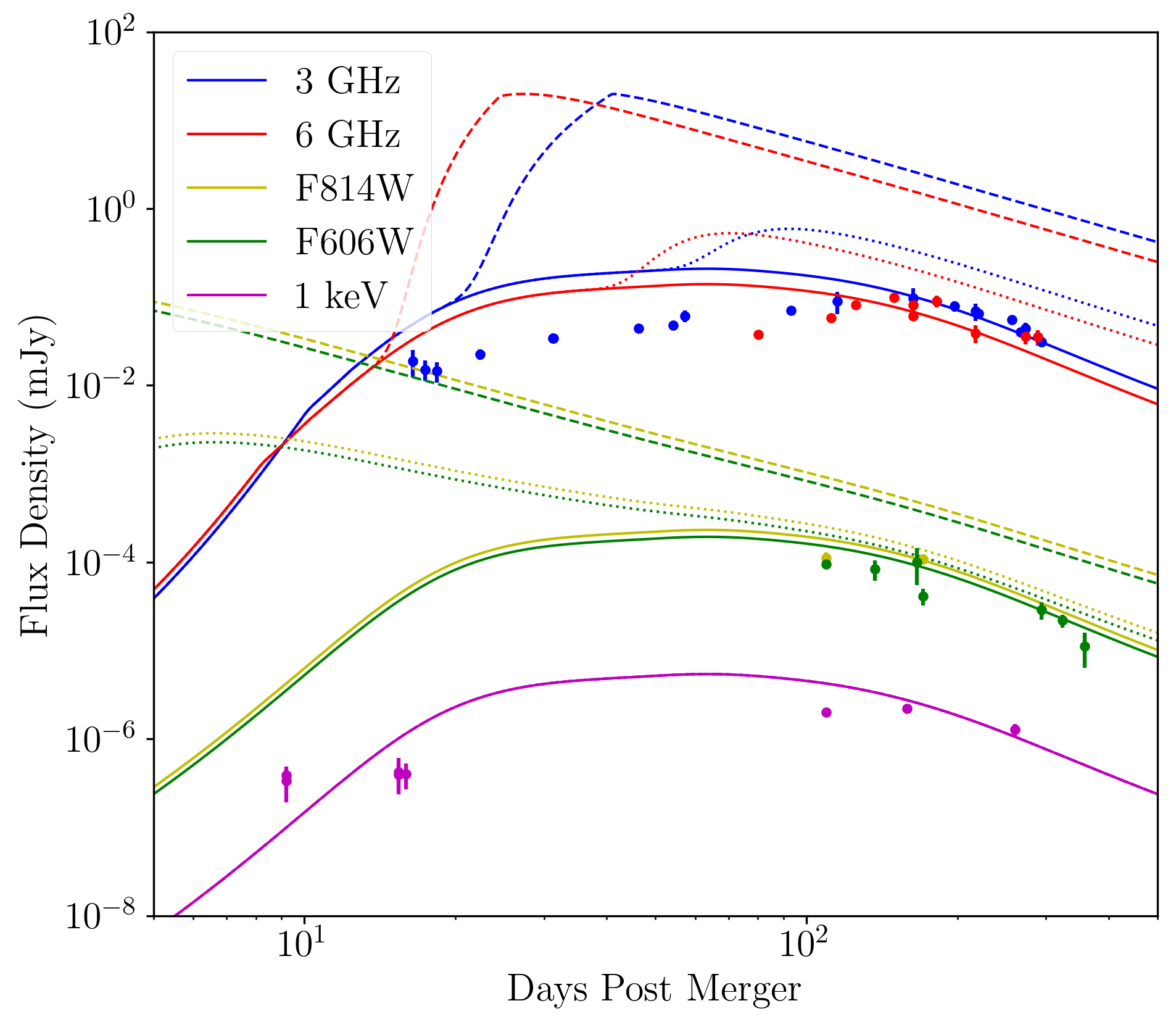}
    \caption{The GRB afterglow of GRB170817A in optical, x-ray, and radio \citep{Evans2017, Hallinan2017, Haggard2017, Mooley2018} and a Gaussian jet light curve model using the median parameters from \citet{Lamb2019_170817}.  PWNe from magnetars with $L_0 = 10^{50}$ ($E_{\rm rot} = 10^{52}$) and $L_0 = 10^{49}$ ($E_{\rm rot} = 10^{51}$) are shown with dashed and dotted lines, respectively, while the afterglow with no PWN is shown with solid lines.  The other PWN parameters are taken from the fiducial KN/SGRB case.}
    \label{fig:170817}
\end{figure}

The SGRB counterpart to the binary neutron star (BNS) merger GW170817, GRB170817A, is the closest SGRB to date \citep{Fong2022, Nugent2022} and one of the most studied transients in history \citep[e.g.,][]{Abbott2017mm, Abbott2017gwgrb, Abbott2017kn, Margutti2021}.  The total mass of the BNS system was inferred to be $\sim$ 2.7 $M_\odot$ \citep{Abbott2019prop} with $<$ 0.1 $M_\odot$ of ejected material, which is larger than the inferred maximum non-rotating neutron star mass $M_{\rm TOV}$ of 2.2 -- 2.3 $M_\odot$.  However, several studies propose that GW170817 could have had a long-lived or infinitely stable neutron star, especially if the spin-down luminosity is dominated by gravitational wave emission \citep{Ai2018, Yu2018, Piro2019, Sarin2021_review, DuPont2024}.  Figure \ref{fig:170817} shows the afterglow of GRB170817A in optical, x-ray, and radio \citep{Evans2017, Hallinan2017, Haggard2017, Mooley2018, Lyman2018}, along with a Gaussian jet light curve model using the median parameters from \citet{Lamb2019_170817}.  We note that the afterglow models used here and in \citet{Lamb2019_170817} have slightly different physics\footnote{Our model has updated versions of jet spreading and SSA compared to \citet{Lamb2019_170817}.  Our use of median parameters instead of the most likely parameters will also cause some discrepancy between our light curve and the data.}, and thus our model does not match the data exactly.  We do not model the emission from a kilonova component here, and the optical opacity for absorption of PWN emission is taken as $\kappa$ = 1 cm$^2$ g$^{-1}$, which is expected for the lanthanide-free ejecta from mergers with long-lived neutron star remnants \citep{Metzger2014knrb, Lippuner2017, Tanaka2020}.  We show a GRB afterglow without a PWN and two with initial magnetar luminosities of $L_0 = 10^{50}$ ($E_{\rm rot} = 10^{52}$) and $L_0 = 10^{49}$ ($E_{\rm rot} = 10^{51}$), with all other parameters being taken from the fiducial case. With a PWN, both the optical and radio emission are significantly more luminous than observed, and the radio emission shows a distinctive rise at around 10 -- 100 days, would would have been readily observable.  The optical emission would have also risen on the timescale of 1 -- 2 days and would have been detectable in the initial kilonova emission.  This shows that if the remnant of the 170817 BNS merger is a stable neutron star, the rotational energy of the neutron star emitted in electromagnetic radiation would have to be $\lesssim 10^{51}$ erg in order to not be detectable in the afterglow, which gives a conservative lower limit on the initial spin period of $\sim$ 7 ms for a 2.2 $M_\odot$ neutron star, similar to what was found by \citet{Murase2018}.  A remnant neutron star with initial spin period of $<$ 7 ms would also have increased the kilonova bolometric luminosity by 1 -- 3 orders of magnitude \citep{Sarin2022} depending on the exact spin period, which would have been detectable for spin periods $\sim$ 1 ms but may not be for spin periods $\sim$ 5 ms due to uncertainties in kilonova modeling \citep{Sarin2024kn, Brethauer2024, Pognan2025}.

A recent case is GRB210702A, which showed a frequency-dependent rebrightening as predicted by the PWN model \citep{deWet2024}.  However, GRB210702A is an LGRB at $z = 1.160$ \citep{Xu2021} with the rebrightening starting at $\sim$ 10 days in the observer frame, bringing the rest frame rebrightening time to $\sim$ 5 days.  Given the timescale from Equation \ref{eqn:tesc_ff}, this would require $\lesssim$ 0.01 $M_\odot$ of ejecta traveling at a significant fraction of the speed of light, which is unrealistic for even a kilonova.  The distance is also far outside the expected horizon for our fiducial parameters, so the effect of the PWN would be difficult to observe in this object.  Thus, it seems unlikely that the source of the rebrightening could be a PWN.

\section{Summary} \label{sec:sum}

Non-thermal light curves from magnetar-powered GRBs can show emission from the GRB afterglow, PWN, and ejecta afterglow.  The timescales for the peak of each emission component are different in both radio and x-ray for our fiducial parameters, making each component potentially distinguishable in the resulting light curve.  The GRB afterglow will peak first, followed by the PWN, then the ejecta afterglow.  The GRB and ejecta afterglow peak timescales are set by the deceleration timescale and the movement of the spectral peak across the band, while the PWN peak timescale is set by absorption processes within the transient ejecta, giving different behaviours across different bands for the peak timescales.

We show the light curves for fiducial SNe/LGRBs and KNe/SGRBs, both on- and off-axis, at 1 GHz, 100 GHz, 1 keV, and 100 keV.  At 1 GHz, the PWN dominates the emission at $\sim$ 6 years for SNe/LGRBs and $\sim$ 100 days for KN/SGRBs, although PWN can be subdominant to a KN afterglow if the density of the surrounding medium is high enough.  At 100 GHz, the PWN dominates the emission at $\sim$ 1 year for SNe/LGRBs and $\sim$ 15 days for KN/SGRBs, regardless of the density of the ambient medium.  The PWN will never be detectable at 1 keV due to strong absorption, and while the PWN does dominate the 100 keV emission over the first few decades, the emission is faint enough that detecting it with current instruments is impossible.  

We simulate our fiducial KN/SGRB observed at 1 and 100 GHz and fit it with a three-component model using a prior that can effectively turn off the PWN component.  The posterior shows a 0$\%$ probability that the transient can be explained without a PWN component.  The two distinguishing features of the PWN component are the frequency-dependent rise and the fast rise from only afterglow emission to PWN peak.  The ejecta afterglow can show a frequency-dependent rise time in atypical cases, but the rise will slow down in those cases, making it possible to distinguish through high-cadence observations.

The detection horizon for the fiducial PWN in these scenarios is $z \sim 0.05 - 0.06$ for SN/LGRBs and $z \sim 0.2 - 0.3$ for KN/SGRBs with current instruments and $z \sim 0.2 - 0.3$ for SN/LGRBs and $z \sim 0.7 - 1.5$ for KN/SGRBs with next-generation instruments.  The PWN emission would be detectable in optical, but will likely compete with the thermal emission from the transient.  Our simple treatment of the emission components neglects asymmetry and interaction between the different components, which can have a strong effect but is difficult to describe in a self-consistent way.  We modeled the potential non-thermal emission from GRB170817A and GRB210702A and found that neither of them are likely to contain a PWN.  The optimal observing strategy for detecting PWNe in GRB afterglows is multiband, high-cadence radio follow-up from 10 -- 100 days of nearby SGRBs with an x-ray plateau or extended emission.

\section*{Acknowledgements}

The authors thank Simon de Wet for sharing the data for GRB210702A, Shiho Kobayashi and Cairns Turnbull for helpful discussions, and the anonymous referee for their helpful comments.  C. M. B. O. and G. P. L. acknowledge support from the Royal Society (grant Nos. DHF-R1-221175 and DHF-ERE-221005).  N. S. acknowledges support from the Knut and Alice Wallenberg Foundation through the "Gravity Meets Light" project and by the research environment grant ``Gravitational Radiation and Electromagnetic Astrophysical Transients'' (GREAT) funded by the Swedish Research Council (VR) under Dnr 2016-06012.

\section*{Data Availability}

The models are available for public use within {\sc{Redback}}~\citep{Sarin_redback}.



\bibliographystyle{mnras}
\bibliography{ref} 




\appendix

\section{Overview of the PWN Model} \label{sec:pwn}

This model is based on the analytic scalings presented in \citet{Murase2021}.  The dynamics for the model are calculated in the same way as \citet{Sarin2022} and \citet{Omand2024}; see those papers for details. 

\subsection{Emission}

The spin-down luminosity of the PWN can be given by the equation \citep{Lasky2017}
\begin{equation}
    L_{\rm SD}(t) = L_0 \left( 1 + \frac{t}{t_{\rm SD}} \right)^{\frac{1+n}{1-n}},
    \label{eqn:llasky}
\end{equation}
where $L_0$ is the initial PWN luminosity, $t_{\rm SD}$ is the spin-down time, and $n$ is the pulsar breaking index.  Equation \ref{eqn:llasky} can be substituted for other spin-down formalisms, such as in \citet{Sarin2022}.  The magnetic field of the PWN is estimated as
\begin{equation}
    B_{\rm neb} \approx \sqrt{\frac{6 \epsilon_{\rm B}}{R_{\rm ej}^3} \int L_{\rm SD} dt},
    \label{eqn:bneb}
\end{equation}
where $\epsilon_{\rm B}$ is the fraction of spin-down energy carried in the magnetic field of the PWN.  This value is usually around 0.003 for Galactic PWNe \citep{Kennel1984, Tanaka2010, Tanaka2013}, but has been inferred to be lower in some superluminous supernovae \citep{Vurm2021}.  The characteristic synchrotron frequency, where $\nu F_\nu$ peaks, is
\begin{equation}
    \nu_b \approx \frac{3}{4\pi}\gamma_b^2 \frac{eB_{\rm neb}}{m_ec},
    \label{eqn:nub}
\end{equation}
where $\gamma_b$ is the electron injection Lorentz factor, usually taken to be $10^4-10^7$ for Galactic PWNe \citep{Tanaka2013}.  This is true if $\nu_b$ is lower than the maximum synchrotron frequency $\nu_M \sim 3.8 \times 10^{22}$ Hz. $F_\nu$ in the fast cooling limit is
\begin{equation}
    F_\nu = F_{\nu 0} \left( \frac{\nu}{\nu_0} \right)^{1-\beta_l} = \frac{\epsilon_e L_{\rm SD}}{8\pi d^2 \nu_0 \mathcal{R}_0(1 + Y_{\rm IC})} \left( \frac{\nu}{\nu_0} \right)^{1-\beta_l}, 
    \label{eqn:fnu}
\end{equation}
for $l = 1 (2)$ when $\nu < (>) \nu_0$, where $\nu_0$ = min$[\nu_b, \nu_M]$ is the peak of the $\nu F_\nu$ spectrum, $\epsilon_e \approx 1 - \epsilon_B$ is the fraction of spin-down energy carried by non-thermal leptons, 
\begin{equation}
    \mathcal{R}_0 = \frac{1}{2 - q_1} - \frac{1}{2 - q_2}
    \label{eqn:bolocorr}
\end{equation}
is the bolometric normalization factor, $Y_{\rm IC}$ is the Compton $Y$ parameter (which we assume to be 0, since we only model a synchrotron nebula), and $d$ is the distance to the source.   The photon indices $\beta_1$ and $\beta_2$ introduced by $F_\nu \propto \nu^{1 - \beta_n}$ are $\beta_1 =$ max$[3/2,(2+q_1)/2]$ and $\beta_2 = (2+q_2)/2$, where $q_1 < 2$ and $q_2 > 2$ are the low and high energy spectral indices of the non-thermal leptons. 

\subsection{Absorption}

PWN emission can be subject to various absorption processes depending on the frequency of the emission.  At radio frequencies, synchrotron self-absorption (SSA) and free-free absorption (FFA) are the main processes.  At 10 eV -- 10 keV bands, the dominant process is photoelectric (bound-free) absorption. At 10 keV -- 10 MeV bands, the dominant process is Compton scattering.  Other bands have different processes, such as Bethe-Heitler pair production for $>$ 10 MeV bands and bound-bound absorption in the optical/infrared bands \citep{Kashiyama2016}, but we only include absorption processes in the radio and 10 eV -- 10 MeV bands.

The SSA frequency can be estimated by \citep{Murase2021}
\begin{equation}
    \pi \frac{R_{\rm ej}^2}{d^2}2kT_{\rm ssa} \frac{\nu_{\rm ssa}^2}{c^2} = F_{\nu 0} \left( \frac{\nu_{\rm ssa}}{\nu_0} \right)^{1-\beta_1},
    \label{eqn:nusa1}
\end{equation}
where
\begin{equation}
    T_{\rm ssa} = \frac{1}{3k} \left( \frac{4\pi m_e c \nu_{\rm ssa}}{3eB_{\rm neb} }\right)^{1/2} m_ec^2
    \label{eqn:tsa}
\end{equation}
is the brightness temperature at $\nu_{\rm ssa}$.  This approximately gives
\begin{equation}
    \nu_{\rm ssa} \approx \left( \frac{3^{3/2}e^{1/2} B_{\rm neb}^{1/2} F_{\nu 0} \nu_0^{\beta_1-1} d^2}{4\pi^{3/2}m_e^{3/2}c^{1/2}R_{\rm ej}^2}
    \right)^{\frac{2}{2\beta_1+3}} .
    \label{eqn:nusa2}
\end{equation}
The optical depth for free-free absorption is \citep{Lang1999, Murase2017}
\begin{equation}
    \tau_{\rm ff} \approx 8.4 \times 10^{-28} n_e^2 R_{\rm ej} \bar{Z}^2 \left( \frac{\nu}{10 \text{ GHz}} \right)^{-2.1} ,
    \label{eqn:tauff}
\end{equation}
where
\begin{equation}
    n_e = \frac{3}{4\pi R_{\rm ej}^3} \frac{M_{\rm ej} Y_{\rm fe}}{m_p}
    \label{eqn:ne}
\end{equation}
is the electron density, $\bar{Z}$ is the average atomic number of the ejected material, and
\begin{equation}
    Y_{\rm fe} \equiv \frac{n_{\rm fe}}{n_p + n_n}
    \label{eqn:Yfe}
\end{equation}
is the free electron fraction, with $n_{\rm fe}$ being the free electron density.

The optical depths for optical and x-ray absorption can be generally written as $\tau = \kappa \rho_{\rm ej} R_{\rm ej}$ for opacity $\kappa$, where the opacity for photoelectric absorption is approximated as \citep{Murase2015, Kashiyama2016}
\begin{equation}
    \kappa_{\rm pe} = 11 \left(\frac{\bar{Z}}{10}\right)^3 \left( \frac{h\nu}{10 \text{ keV}} \right)^{-3} \text{ cm$^2$ g$^{-1}$}
    \label{eqn:kpe}
\end{equation}
and for Compton scattering as \citep{Murase2015}
\begin{equation}
    \kappa_{\rm comp} = \frac{\sigma_{\rm KN} Y_e}{m_p},
    \label{eqn:kcomp}
\end{equation}
where $\sigma_{\rm KN}$ is the Klein-Nishina cross section \citep{Klein1929} and $Y_e$ is the electron fraction of the ejecta.  The optical opacity is generally taken to be free parameter in most light curve models \citep[e.g.][]{Arnett1982, Villar2017kn}, and we adopt the same treatment here.

\subsection{Parameters and Default Priors}

This model is implemented as the {\sc{PWN}} model in {\sc{general$\_$synchrotron$\_$models}} within {\sc{Redback}}.  Along with the parameters that can freely vary during inference, there are a number of parameters that have default values consistent with a supernova, but that can be changed if the user wants to model a kilonova.  These can also be given priors and marginalized over if needed.  The parameter $\kappa_\gamma$ here is only used in the dynamics on the supernova, not in the calculation of the PWN spectrum.  A summary of the parameters and priors is given in Table \ref{tbl:pwnparams}.

\begin{table}
\centering
\begin{tabular}{cccc}
   Parameter & Definition & Units & Prior/Value \\ \hline
   $M_{\rm ej}$ & Ejecta Mass & $M_{\odot}$ & L[0.1, 100] \\
   $L_0$ & Initial Spin-Down Luminosity & erg s$^{-1}$ & L[$10^{40}$, $10^{50}$]  \\
   $t_{\rm SD}$ & Spin-Down Time & s & L[$10^{2}$, $10^{8}$] \\
   $n$ & Magnetar Braking Index &  & U[1.5, 10]  \\
   $\epsilon_{\rm B}$ & Magnetic Field Partition Parameter & & L[$10^{-7}$, $1$] \\
   $\gamma_b$ & Electron Injection Lorentz Factor & & L[$10^{2}$, $10^{7}$] \\ \hline
   $E_{\rm K}$ & Initial Kinetic Energy & erg & $10^{51}$  \\
   $\kappa$ & Ejecta Optical Opacity & cm$^2$ g$^{-1}$ & 0.1  \\
   $\kappa_\gamma$ & Ejecta Gamma-Ray Opacity & cm$^2$ g$^{-1}$ & 0.01 \\
   $q_1$ & Low Energy Spectral Index & & 1.5 \\
   $q_2$ & High Energy Spectral Index & & 2.5 \\
   $\bar{Z}$ & Ejecta Average Atomic Number & & 8 \\
   $Y_e$ &  Ejecta Electron Fraction & & 0.5 \\
   $Y_{\rm fe}$ &  Ejecta Free Electron Fraction & & 0.0625 \\
\end{tabular}
\caption{The parameters and default priors implemented for this model. Priors are either uniform (U) or log-uniform (L). }
\label{tbl:pwnparams}
\end{table}

\section{Posterior for Inference on a Simulated Light Curve}

\begin{figure*}
    \centering
    \includegraphics[width=\linewidth]{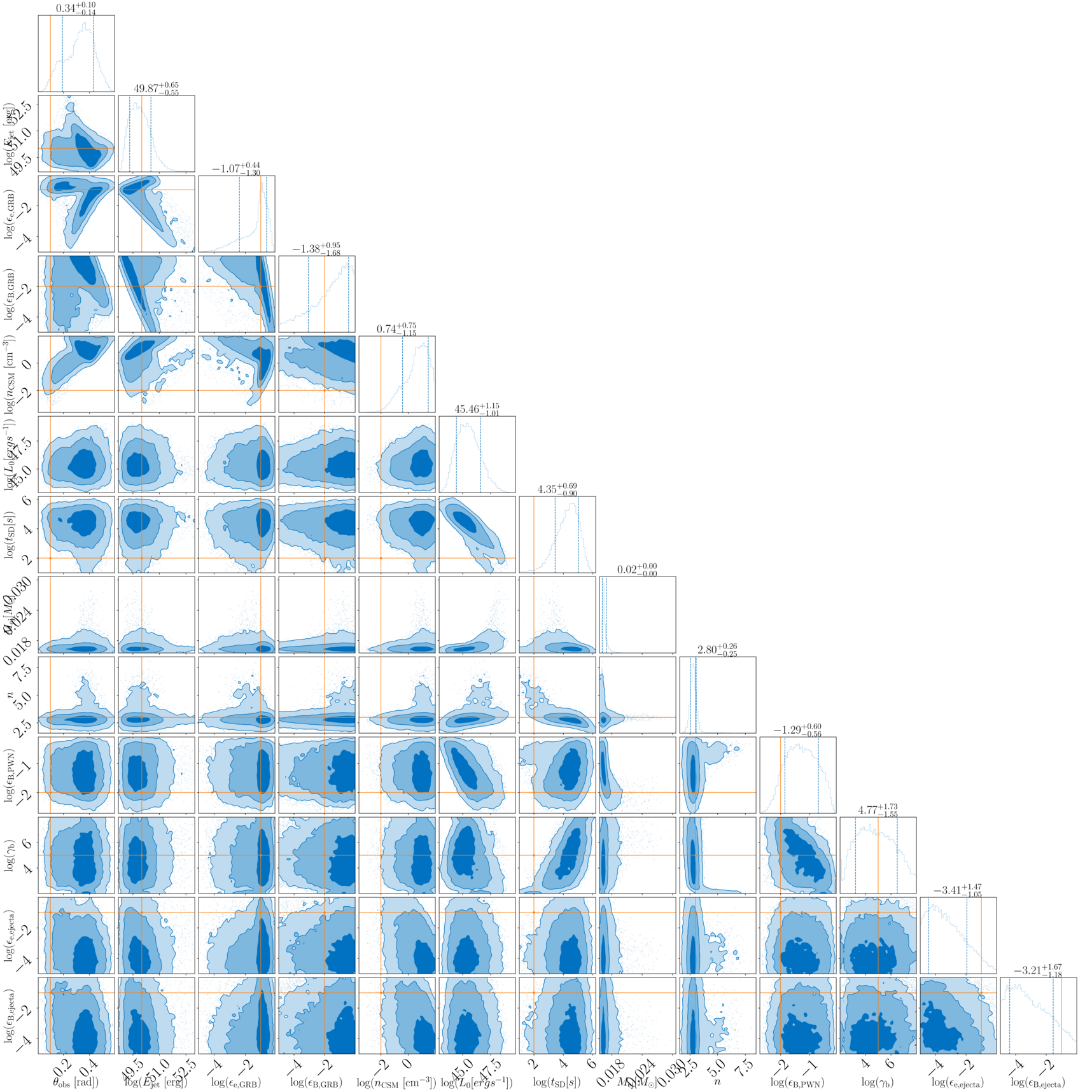}
    \caption{The parameter posterior inferred for the simulated kilonova/SGRB.  The orange dots and lines indicate the injected parameters.}
    \label{fig:sim_corner}    
\end{figure*}

The posterior for the fit done in Section \ref{sec:inf} is shown in Figure \ref{fig:sim_corner}.

\bsp	
\label{lastpage}
\end{document}